\documentclass[journal]{IEEEtran}
\pdfoutput=1
\ifCLASSINFOpdf
\else
\fi

\usepackage{cite}
\usepackage{booktabs}
\usepackage{multirow}
\usepackage{amsmath,amssymb,amsfonts,amsthm}
\usepackage{algorithm}
\usepackage{algorithmic}
\newtheorem{theorem}{Theorem}  
  
\usepackage{textcomp}
\usepackage{url}
\usepackage{bbm}
\usepackage{enumerate}

\usepackage{xcolor}
\usepackage{graphicx}
\usepackage{graphics}
\usepackage{subfigure}

\begin{document}

\title{CoEdge: Cooperative DNN Inference with \\Adaptive Workload Partitioning \\over Heterogeneous Edge Devices}

\author{Liekang~Zeng,~\IEEEmembership{Student Member,~IEEE,}
        Xu~Chen,~\IEEEmembership{Senior Member,~IEEE,}
        Zhi~Zhou,~\IEEEmembership{Member,~IEEE,}
        Lei~Yang,~\IEEEmembership{Senior Member,~IEEE}
        and~Junshan~Zhang,~\IEEEmembership{Fellow,~IEEE}
\thanks{L. Zeng, X. Chen and Z. Zhou are with the School of Computer Science and Engineering, Sun Yat-sen University, Guangzhou, Guangdong, 510006 China (e-mail: zenglk3@mail2.sysu.edu.cn, chenxu35@mail.sysu.edu.cn, zhouzhi9@mail.sysu.edu.cn).}
\thanks{L. Yang is with the Department of Computer Science and Engineering, University of Nevada, Reno, NV, 89557 USA (e-mail: leiy@unr.edu).}
\thanks{J. Zhang is with the School of Electrical, Computer and Energy Engineering, Arizona State University, Tempe, 85287 USA (e-mail: junshan.zhang@asu.edu).}
}

\maketitle

\begin{abstract}
Recent advances in artificial intelligence have driven increasing intelligent applications at the network edge, such as smart home, smart factory, and smart city.
To deploy computationally intensive Deep Neural Networks (DNNs) on resource-constrained edge devices, traditional approaches have relied on either offloading workload to the remote cloud or optimizing computation at the end device locally.
However, the cloud-assisted approaches suffer from the unreliable and delay-significant wide-area network, and the local computing approaches are limited by the constrained computing capability.
Towards high-performance edge intelligence, the cooperative execution mechanism offers a new paradigm, which has attracted growing research interest recently.
In this paper, we propose CoEdge, a distributed DNN computing system that orchestrates cooperative DNN inference over heterogeneous edge devices.
CoEdge utilizes available computation and communication resources at the edge and dynamically partitions the DNN inference workload adaptive to devices' computing capabilities and network conditions.
Experimental evaluations based on a realistic prototype show that CoEdge outperforms status-quo approaches in saving energy with close inference latency, achieving up to 25.5\%$\sim$66.9\% energy reduction for four widely-adopted CNN models.
\end{abstract}


\IEEEpeerreviewmaketitle

\section{Introduction}\label{sec:introduction}

\IEEEPARstart{R}{ECENT} years have witnessed an ever-increasing number of Internet of Things (IoT) devices diving into miscellaneous application domains, e.g., smart home \cite{stojkoska2017review}, smart factory \cite{shrouf2014smart}, autonomous driving \cite{gerla2014internet}, etc.
This trend also drives the community to build smarter, faster, and greener intelligent applications at the network edge, pushing remarkable progress in smart healthcare, security inspection and disease detection \cite{rahmani2015smart, shi2020carpool, acharya2017deep}.
Meanwhile, advances in Deep Neural Networks (DNNs) have shown unprecedented ability in learning abstract representation and extracting high-level features, promoting significant improvement in processing human-centric contents \cite{sze2017efficient}.
Motivated by this success, it is envisioned that employing DNNs to edge devices would enable and boost intelligent services, supporting brand new smart interactions between humans and their physical surroundings.

The essential demand of these services is to respond user's queries timely, e.g., recognizing voice commands \cite{deng2013new}, inspecting visitor's faces \cite{simonyan2014very}, and detecting heartbeat frequency \cite{acharya2017deep}, all within a matter of milliseconds.
This also implies a soft-realtime requirement - if the result comes late, the user may turn to other applications, and the result can even be out of date and meaningless.
Therefore, minimizing response latency and promising users' experience is of paramount importance.
However, DNN-based applications are typically computation-intensive and resource-hungry \cite{sze2017efficient}, and service providers traditionally appeal to the resource-abundant cloud to satisfy the stringent responsiveness requirement \cite{li2019edge}.
Yet the Quality-of-Service (QoS) can still be poor and unsatisfactory due to the unreliable and delay-significant wide-area network connection between edge devices and the remote cloud \cite{ouyang2018follow, chen2018thriftyedge}.
What's worse, for many smart applications with human in the loop, the sensory data can contain highly sensitive or private information. 
Offloading these data to the remote datacenter owned by curious commercial companies inevitably raises users' privacy concerns.

\begin{figure}[t]
  \centering
  \includegraphics[width=0.9\linewidth]{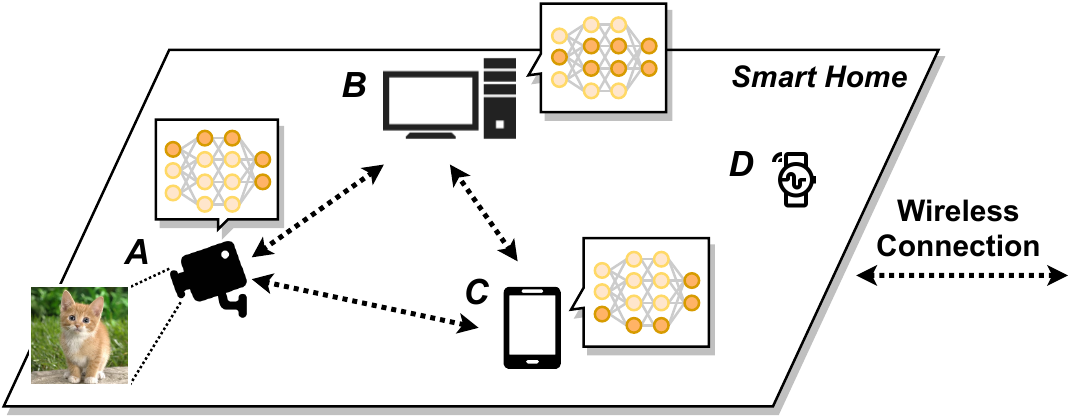}
  \caption{An example of cooperative DNN inference in a smart home scenario. As the raw image is captured, device \textit{A} decides a cooperative execution plan and distributes the workload to devices \textit{B} and \textit{C}. According to the plan, the devices perform cooperative inference in response to the DNN task.}
  \label{fig:scenario}
\end{figure}

\begin{figure*}[t]
  \centering
  \includegraphics[width=0.8\textwidth]{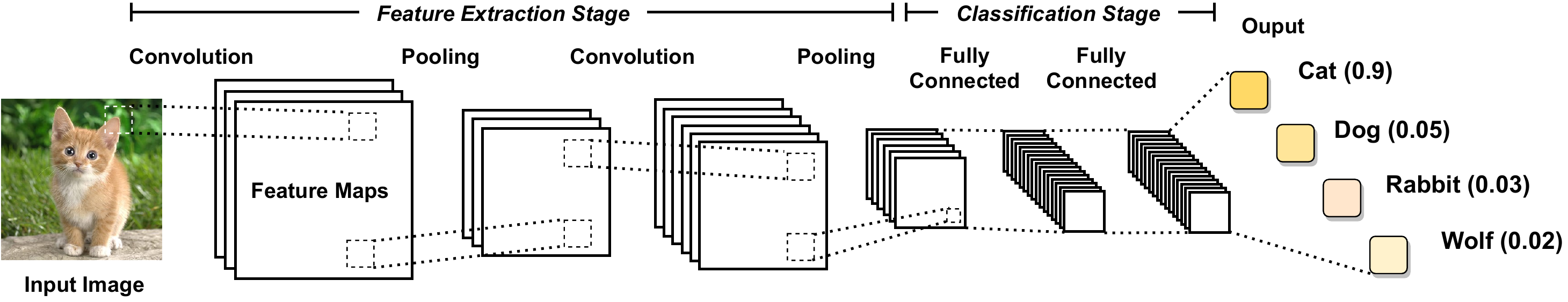}
  \caption{Conventional CNN inference workflow, which is typically in two stages. In the first stage, CNN processes the input image to extract hidden features through operations like convolution and pooling, and generates multidimensional feature maps. In the second stage, CNN classifies the feature maps by fully-connected layers and obtains the inference result.}
  \label{fig:DNN}
\end{figure*}

Intuitively, keeping data locally and processing tasks without external remote assists will preserve user privacy and avoid the remote network transmission. 
Unfortunately, local edge devices are generally with limited computing capability, making it hard to fulfill DNN execution under the latency Service-Level-Objective (SLO).
For instance, if a smart home camera runs CNN-based face recognition to provide real-time inspection and warning, the response delay when running DNN locally may last for a few seconds, resulting in poor user experience and completely unusable service.

To tackle these challenges, a promising approach is to exploit available computation resources in the proximity to the data source with the emerging edge intelligence paradigm \cite{zhou2019paving}.
Instead of uploading data to the remote cloud or keeping all computation at the single local device, edge intelligence enjoys real-time response as well as privacy preservation by offloading computing workload within a manageable range.
As Fig. \ref{fig:scenario} illustrates, we can utilize the diverse computing resources in a smart home (with inspection camera, smartphone, tablet, and desktop PC) to accelerate the CNN-based face recognition.
Specifically, the source device \textit{A} can distribute the inference workload to devices \textit{B} and \textit{C}, and perform cooperative inference via high-bandwidth local wireless connection (e.g., WiFi).
Nevertheless, this paradigm brings some key challenges to be addressed: (1) how to decide the workload assignment tailored to the resource heterogeneity of edge devices, (2) how to optimize the system performance with the presence of network dynamics, and (3) how to orchestrate computation and communication during cooperative inference runtime.

To answer these questions, we propose CoEdge (\textbf{Co}operative \textbf{Edge}), a runtime system that orchestrates cooperative DNN inference over multiple heterogeneous edge devices.
CoEdge does not apply any structural modifications or tuning requirements to the given DNN model, and does not sacrifice model accuracy as it reserves input data and model parameters of the given DNN model.
CoEdge employs a similar parallel workflow as DeepThings \cite{zhao2018deepthings}, where the input is split initially and the execution is parallelized on multiple devices at runtime.
While DeepThings leverages a layer fusion technique to reduce communication overhead, CoEdge proposes to optimize workload allocation to maximally utilize heterogeneous edge resources.
By optimizing the computation-communication tradeoff, CoEdge optimally partitions the input inference workload, where the partitions' sizes are chosen to match devices' computing capabilities and network conditions to improve system performance in both latency and energy metrics.
We implement CoEdge using a realistic prototype with Raspberry Pi 3, Jetson TX2, and desktop PC.
Experimental evaluations show 7.21$\times$$\sim$4.49$\times$ latency speedup over the local approach and up to 25.5\%$\sim$66.9\% energy saving comparing with existing approaches for four popular DNN models.

In summary, this paper makes the following contributions.
\begin{itemize}
    \item We propose CoEdge, a distributed DNN computing system that orchestrates cooperative inference over heterogeneous devices to minimize system energy consumption while promising response latency requirement.
    \item We identify the impacts of workload partitioning on cooperative inference workflow, and build a constrained programming model on workload distribution optimization. We prove the NP-hardness of the problem, and devise a fast approximated algorithm to decide the efficient partitioning policy in real-time tailored to devices' diverse computing capabilities and network conditions.
    \item We implement a multi-device prototype using heterogeneous edge devices, and evaluate CoEdge on four widely-adopted DNN models to corroborate its superior performance.
\end{itemize}

The rest of this paper is organized as follows. 
Section \ref{sec:background} briefly reviews background on DNN inference, and discusses opportunities and challenges based on a case of cooperative inference.
Section \ref{sec:system_model} presents CoEdge design and its workflow.
Section \ref{sec:formulation} builds the system model and describes our workload partitioning algorithm.
We explain implementation details in Section \ref{sec:implementation} and evaluate the prototype in Section \ref{sec:evaluation}.
Section \ref{sec:related_work} provides related works.
Section \ref{sec:discussion} discusses limitation and extension of CoEdge, and Section \ref{sec:concolusion} concludes.
The appendix (in the supplementary material) details the proofs of Theorem 1 and 2.

\section{Background and Motivation}\label{sec:background}

In this section, we briefly review conventional CNN inference and cooperative inference.
We study a case of cooperative inference and discuss potential challenges behind that.

\subsection{Deep Neural Network Inference}

In this paper, we focus on the classical Convolutional Neural Networks (CNNs) as they are widely adopted across a board spectrum of intelligent services, including image classification, object detection, and semantic segmentation, etc.

Fig. \ref{fig:DNN} depicts a conventional CNN inference for image classification task from a perspective of feature maps.
As we can see, a conventional CNN inference can be viewed as a series of successive algorithmic operations on feature maps.
These operations\footnote{For ease of illustration, only some of the operations are drawn in Fig. \ref{fig:DNN}} comprise of convolution, pooling, batch normalization, activation, and fully-connected computation, etc.
In light of the functionality of the operations, the inference process can be separated into two stages.
The first stage is the feature extraction stage, where the model processes every pixel in the input image to generate hidden feature representations.
Following that, at the second stage, these features are classified by the fully-connected layers, exporting results in a probabilistic form.

\subsection{Case Study: Cooperative Inference with Two Devices}

\begin{table}[t]
\caption{Raspberry Pi 3 Specifications \cite{pi2019}}
\label{tab:pi}
\centering
\begin{tabular}{|c|c|c|}
\hline
Hardware & \multicolumn{2}{c|}{Specifications} \\ \hline \hline
CPU      & \multicolumn{2}{c|}{1.2GHz Quad Core ARM Cortex-A53}\\ \hline
Memory   & \multicolumn{2}{c|}{1GB LPDDR2 900MHz }  \\ \hline
GPU      & \multicolumn{2}{c|}{No GPU} \\ \hline \hline
Power    & \begin{tabular}[c]{@{}c@{}}Idle\\ Fully Loaded\\ Average Observed\end{tabular} & \begin{tabular}[c]{@{}c@{}}1.3W\\ 6.5W\\ 3W\end{tabular} \\ \hline
\end{tabular}
\end{table}

\begin{table}[t]
\caption{Jetson TX2 Specifications \cite{jetson2019}}
\label{tab:jetson}
\centering
\begin{tabular}{|c|c|c|}
\hline
Hardware & \multicolumn{2}{c|}{Specifications} \\ \hline \hline
CPU      & \multicolumn{2}{c|}{\begin{tabular}[c]{@{}c@{}}2.0GHz Dual Denver 2 +\\ 2.0GHz Quad Core ARM Cortex-A57\end{tabular}}  \\ \hline
Memory   & \multicolumn{2}{c|}{8GB LPDDR4 1.6GHz} \\ \hline
GPU      & \multicolumn{2}{c|}{Pascal Architecture 256 CUDA Core} \\ \hline \hline
Power    & \begin{tabular}[c]{@{}c@{}}Idle\\ Fully Loaded\\ Average Observed\end{tabular} & \begin{tabular}[c]{@{}c@{}}5W\\ 15W\\ 9.5W\end{tabular} \\ \hline
\end{tabular}
\end{table}

The key impediment of deploying CNN at the network edge lies in the gap between intensive CNN inference computation and the limited computing capability of edge devices.
To bridge this gap, we can utilize the cooperative inference mechanism to exploit available computing resources at the edge.
A straightforward solution of that, for example, is the master-worker paradigm that offloads inference workload to external infrastructure.
To obtain a better understanding of cooperative inference, we use a real hardware testbed to emulate this solution.

\begin{figure}[t]
  \centering
  \includegraphics[width=0.55\linewidth]{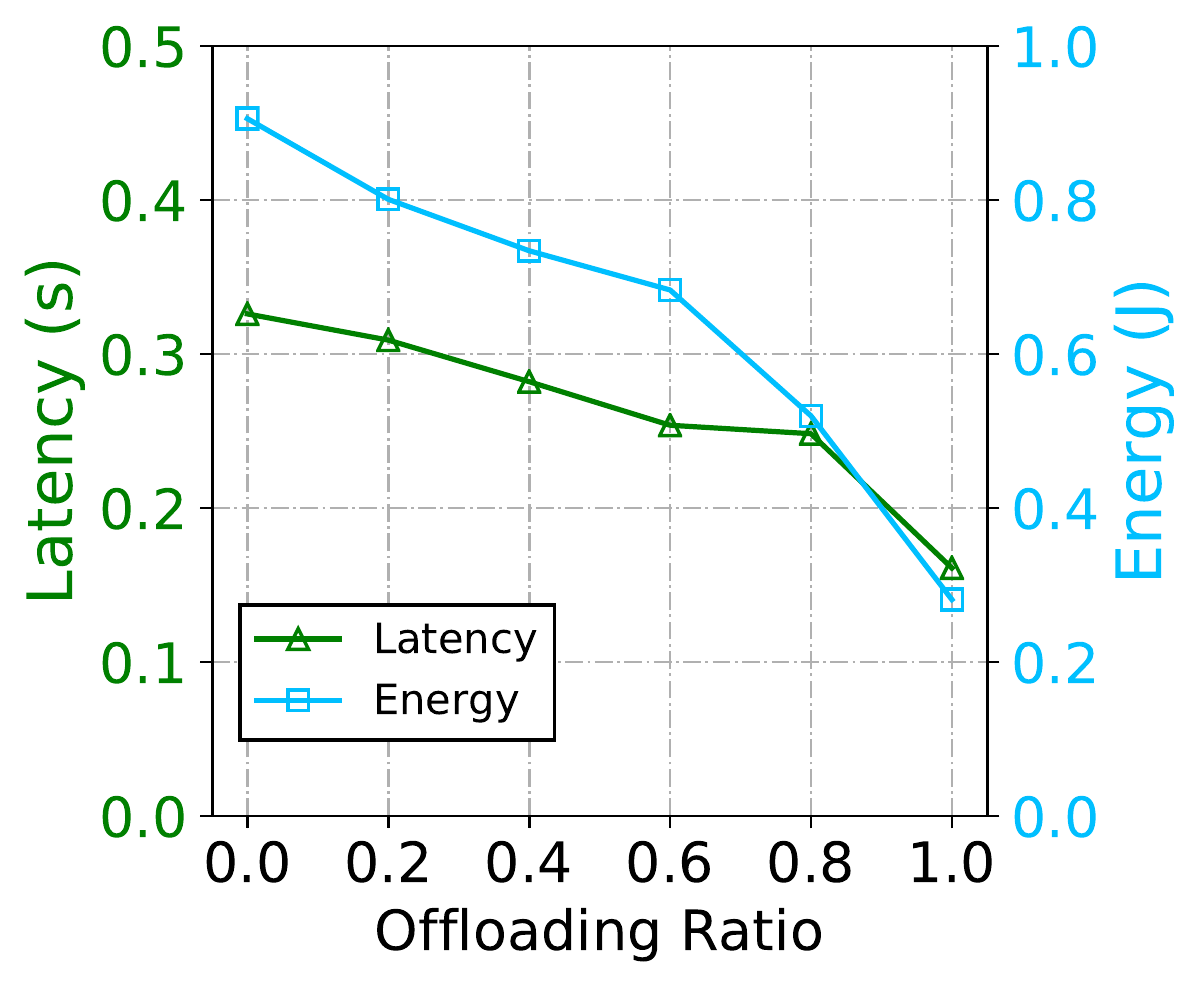}
  \caption{The total latency and energy consumption under varying offloading ratio, i.e., the proportion of data that is offloaded from the Raspberry Pi 3 to the Jetson TX2.}
  \label{fig:offloading_ratio}
\end{figure}

As a case study, we employ a Raspberry Pi 3 and a Jetson TX2, on behalf of weak IoT devices and mobile AI platforms at the edge, respectively.
Table \ref{tab:pi} and \ref{tab:jetson} presents their specifications and reported power parameters, which are measured with Monsoon High Voltage Monitor \cite{monsoon} using the methodology in \cite{hadidi2018distributed}.
For each inference task, we input one single image to the Pi and then offloads a part of the image to the Jetson.
The two devices parallelize the DNN execution and their results are finally aggregated to the Pi as output.
We measure the end-to-end latency of this process, i.e., from the image input to the inference result output; and we record the average latency of fulfilling the inference task over 100 runs.
We implement AlexNet \cite{krizhevsky2012imagenet} with TensorFlow Lite \cite{tensorflow} on both devices, and run the model with the same image from ImageNet \cite{deng2009imagenet}.
For the bandwidth between two devices, we fix it at 1MB/s using the traffic control tool tc \cite{tc2019}.

We define \textit{offloading ratio} to indicate how much data is offloaded from the Raspberry Pi to the Jetson TX2.
For instance, when the ratio is 0.5, we split the input image along the height into two equal parts, and transfer one of them to the Jetson TX2.
In particular, a zero ratio indicates performing inference at the Raspberry Pi locally, while the ratio equals to 1.0 if offloading all workload to the Jetson TX2. 
Fig. \ref{fig:offloading_ratio} shows the latency and energy overheads under varying offloading ratio.
Through this experiment, we derive the following observations.

\begin{itemize}
    \item Jetson TX2 enjoys better performance than Raspberry Pi 3. 
    When the ratio is zero, the system consumption is only the computation cost of Pi, while at the 1.0 case, the total cost comprise of the input offloading overhead, the DNN computation overhead in Jetson and the overhead for transferring result back. 
    However, the former still takes higher cost than the latter in terms of both latency and energy. 
    Note that fully offloading workload to the Jetson (i.e., \textit{offloading ratio} = 1) may not necessarily yield the lowest costs if the network fluctuates.
    \item Cooperative inference is more economical than local inference given the favourable network condition. As the offloading ratio increases, both latency and energy drop. In other words, the system cost decreases via harvesting the cooperator's computing resource.
    \item The curve of latency drops faster as the offloading ratio increases. This is because the DNN execution is parallelized in cooperative inference and the end-to-end inference latency is straggled by the slower one. Therefore, in high bandwidth environments, assigning more workload to Jetson TX2 benefits performance improvement better.
\end{itemize}

\subsection{Merits and Challenges}

We see that, from the above observations, cooperative mechanism has potential to improve inference performance with multiple devices, which are exactly what edge scenarios possess.
More precisely, we envision the deployment of the cooperative inference system in an environment such as smart home or smart factory, wherein the devices are managed by the same owner and thus they are willing to cooperate and share their resources.
This brings several major merits as well as challenges.

\textbf{Merits.} On the one hand, comparing with local inference, cooperative inference has significant potential in reducing latency and energy costs via harvesting idle computing resources at the edge. 
On the other hand, other than the cloud approach that uploads data to the remote datacenter, the cooperative approach keeps data within user's control scope, therefore avoiding the delay-significant wide-area network as well as privacy issues.

\textbf{Challenges.} To effectively exploit computing resources at the edge, we need to felicitously factor the computing capabilities of edge devices considering magnitude and heterogeneity. 
Also, given the dynamic network inherently in edge computing, an efficient workload allocation solution that jointly considers systematic costs is desired. 
More specifically, it is crucial to decide which device to participate in the cooperative inference and how much workload each device affords.
Besides, since the cooperative mechanism parallelizes CNN inference in a distributed manner, the system needs to orchestrate the computation and communication over multiple devices.

We address these challenges by designing a cooperative system, CoEdge, through orchestrating the available resources from heterogeneous edge devices.

\section{CoEdge Design and Workflow}
\label{sec:system_model}

In this section, we present CoEdge design and the workflow of cooperative CNN inference.
We further explore how workload partitioning impacts parallel processing in terms of computation and communication.

\subsection{CoEdge Design}

For ease of illustration, we differentiate between devices on their roles in the cooperation.
For the device that launches a CNN inference task, we label it as the \textit{master device}, and for the device that joins the cooperation, it is marked as the \textit{worker device}.
The master device is responsible for registering participated devices, generating a feasible workload partitioning plan, and managing the cooperative inference over worker devices.
Note that a device can be the master and the worker at the same time since it can retain CNN workload in situ.

\begin{figure}[t]
  \centering
  \includegraphics[width=0.9\linewidth]{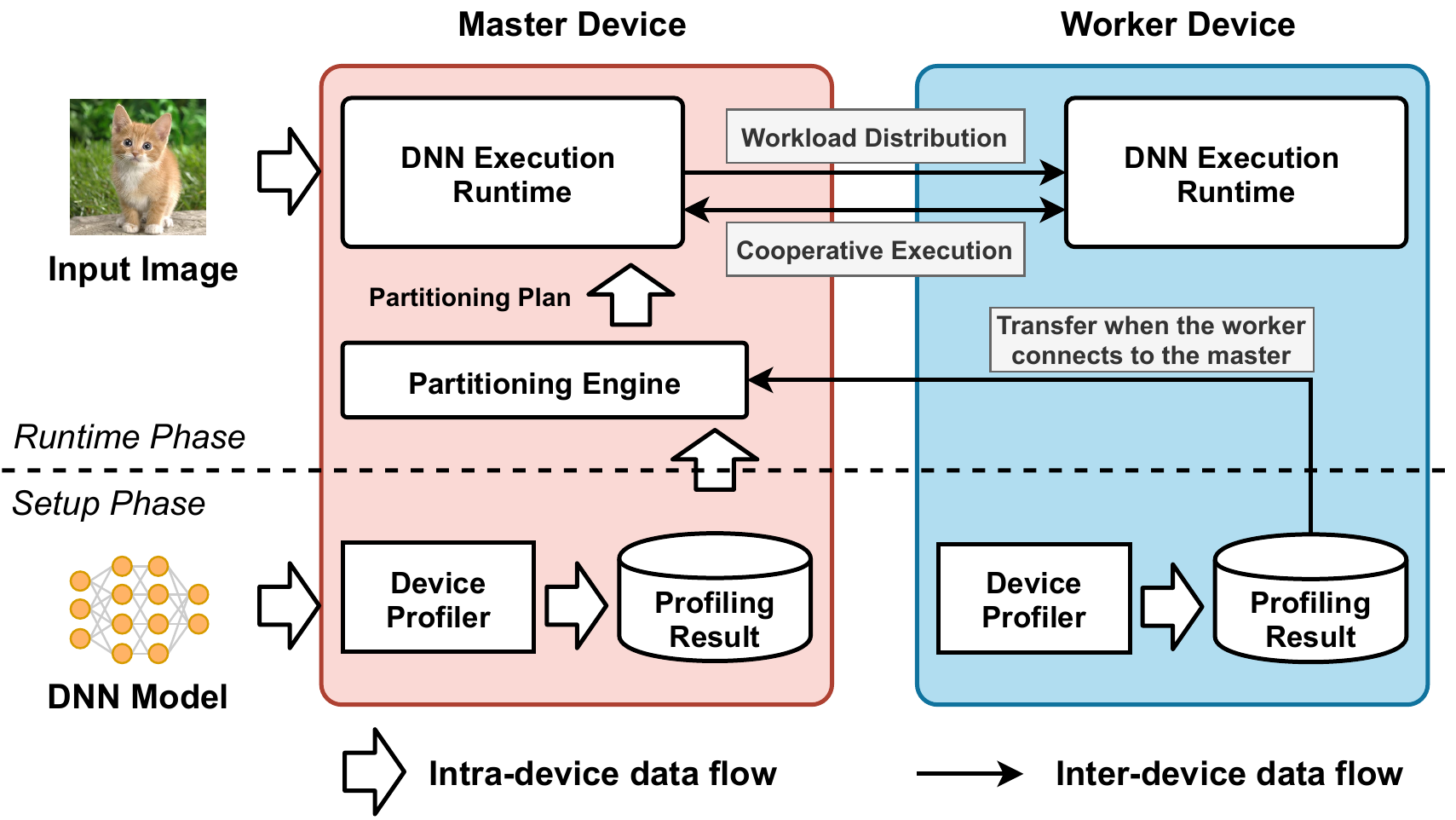}
  \caption{CoEdge architecture overview, which works in two phases. In the setup phase, the devices profile parameters to sketch their computing capabilities information. In the runtime phase, the master device creates a partitioning plan using the collected profiling results. According to the plan, the master device distributes the workload and performs cooperative inference with worker devices. }
  \label{fig:framework}
\end{figure}

Fig. \ref{fig:framework} illustrates the architecture overview of CoEdge, which works in two phases, namely the setup phase and the runtime phase.
In the \textit{setup} phase, CoEdge records the execution profiles of each device.
In the \textit{runtime} phase, CoEdge creates a cooperative inference plan that determines the workload partitions and their corresponding assignment, using the profiling results collected in the setup phase and the network status.
According to the cooperation plan, the master distributes the workload partitions to the workers and then performs cooperative execution collaboratively.

\textbf{Setup phase.} Whenever a CNN-based application is installed, \textit{Device Profiler} runs the CNN models locally and records \textit{Profiling Results}.
These results sketch the device's computing capability, including the computation intensity, computation frequency and power parameters, which will be detailed in Section \ref{sec:modeling}.

\textbf{Runtime phase.} The runtime phase starts when the master raises a CNN inference query. 
As the image inputs, the master establishes connections with worker devices and pulls their profiling results.
Since the size of the profiling results is very small (tens of bytes in our prototype), the transmission overhead for transferring them
is negligible.
As the master receives the profiling data, the \textit{partitioning engine} in the master device generates a workload allocation plan using the adaptive workload partitioning algorithm (explained in Section \ref{sec:algorithm}).
According to the plan, \textit{DNN execution runtime} distributes the workload partitions to workers and performs cooperative inference in response to the query.

\subsection{Cooperative Inference Workflow}

\begin{figure*}[t]
  \centering
  \includegraphics[width=0.8\textwidth]{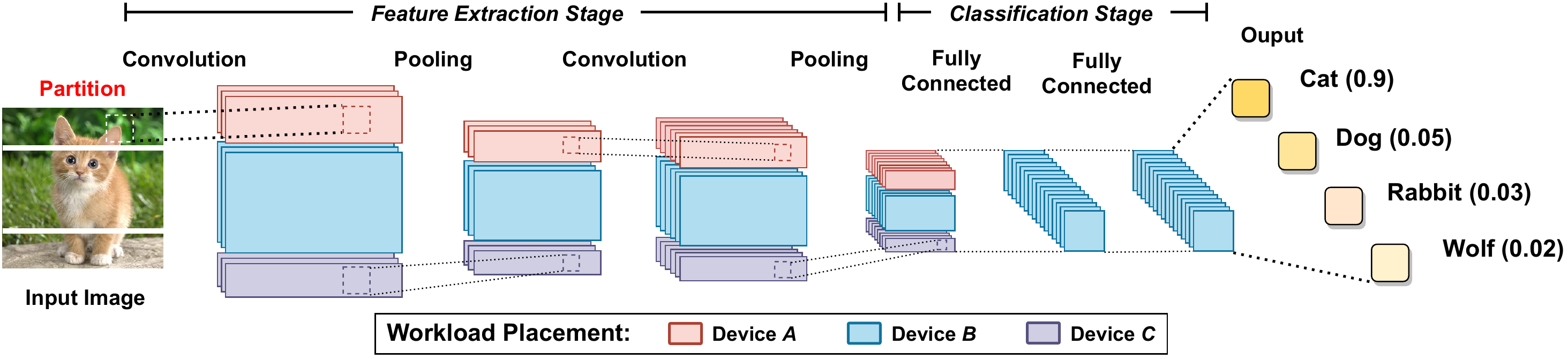}
  \caption{Cooperative CNN inference workflow of CoEdge. The input image is piece-wise partitioned to patches before execution. In feature extraction stage, these patches are distributed to devices \textit{A}, \textit{B} and \textit{C}, respectively, and then in classification stage, the feature map fragments are aggregated to finish the remaining execution.}
  \label{fig:workflow}
\end{figure*}

In this work, we exploit model parallelism to partition CNN inference over multiple devices.
Under model parallelism, CNN model parameters are divided into subsets and assigned to multiple edge nodes.
With respective parameters, each device accepts a necessary part of the input feature maps and generates a portion of the output feature maps.
Concatenating all these portions yields the complete output of each layer.

Fig. \ref{fig:workflow} provides an instance of cooperative inference workflow with three devices from a perspective of feature maps.
The cooperative inference begins when the image is piece-wise split into partitions.
Note that to accommodate devices' heterogeneity, the partition sizes are differentiated to match device capabilities. 
The partitions are then distributed from the master to three devices (i.e., devices \textit{A}, \textit{B}, and \textit{C} in Fig. \ref{fig:workflow}). 
At the feature extraction stage, the three devices execute their workload in parallel, while at the classification stage, their execution results are aggregated to one of them (device \textit{B} in Fig. \ref{fig:workflow}).
This aggregation is to avoid excess communication overhead caused by the nature of fully-connected computation, which requires repeating data access on the feature vectors.

\textbf{Generalization.} Based on the workflow in Fig. \ref{fig:workflow}, it is feasible to accommodate various CNNs with complex structures by redesigning some details.
For example, for CNNs without fully-connected layers (e.g., Network in Network \cite{lin2013network}), we can reduce the classification stage in Fig. \ref{fig:workflow}.
To adapt CNNs with skip connections (e.g., ResNet \cite{he2016deep}), we can keep intermediate output results on each device at the shortcut starting point and release these data at the shortcut destination to collect the data when needed.

\subsection{Impact of Workload Partitioning}
\label{sec:partition}

\begin{figure}[t]
    \centering
    \includegraphics[width=0.8\linewidth]{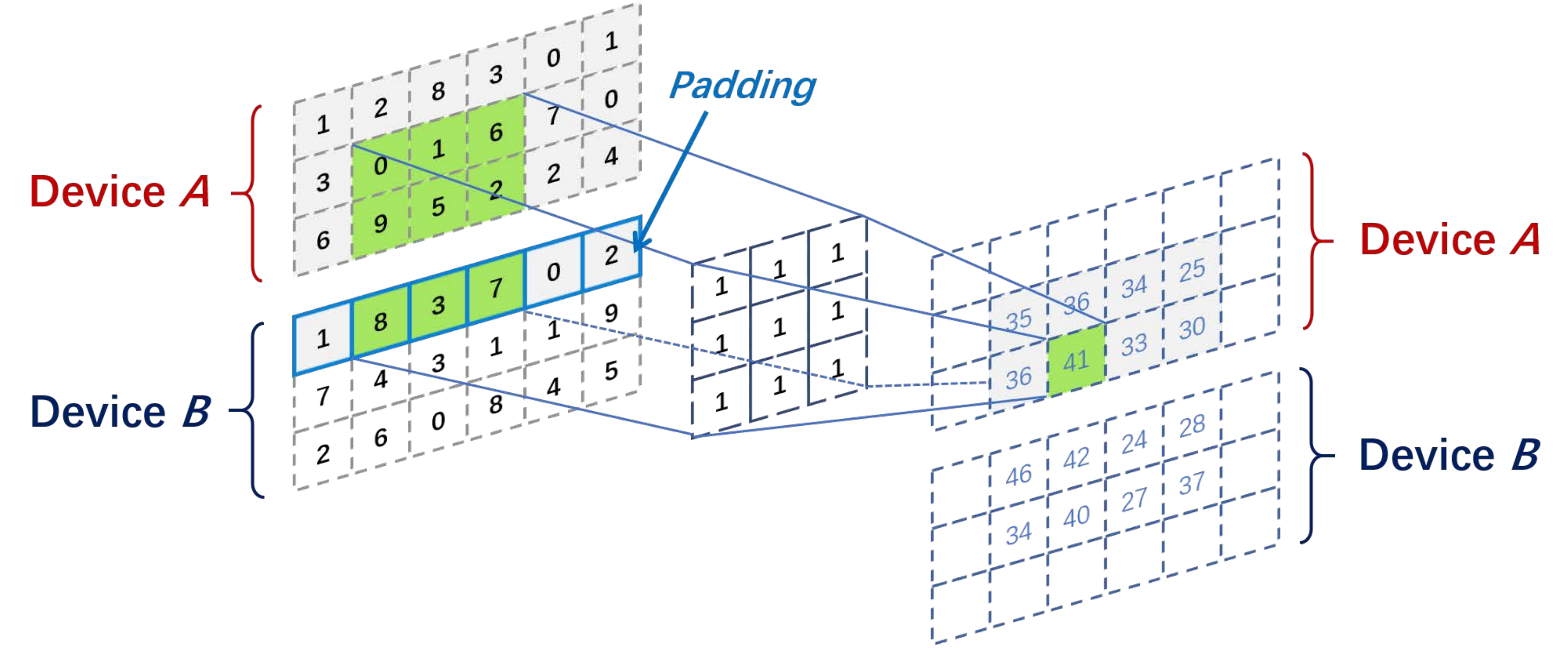}
    \caption{
    Example of a convolution operation for cooperative inference. The input feature map partitions with each of $3\times6$ size locate at devices \textit{A} and \textit{B}, respectively. To generate the output feature map through the $3\times3$ convolution kernel, device \textit{A} needs to pull the padding data of $1\times6$ size from device \textit{B}.
    }
    \label{fig:modeling_partition}
\end{figure}

The way of piece-wise partitioning significantly affects the communication between devices, especially for convolution operations that process data across partition boundaries.
For instance, Fig. \ref{fig:modeling_partition} shows a typical convolution operation with two partitions.
To compute convolution over the $3\times6$ partition with the $3\times3$ kernel, device \textit{A} needs to fetch the $1\times6$ margin row in device \textit{B}'s partition.
In general, for the kernel whose size $k$ is greater than 1, each device needs to pull the padding data of $\lfloor k/2 \rfloor$ size along the split dimension from the neighboring device.
In some extreme cases, when the kernel size is very large but the neighboring partition size is very small, the padding range may even across three or more devices, which could incur extravagant communication overhead.

\begin{figure}[t]
    \centering
    \includegraphics[width=0.85\linewidth]{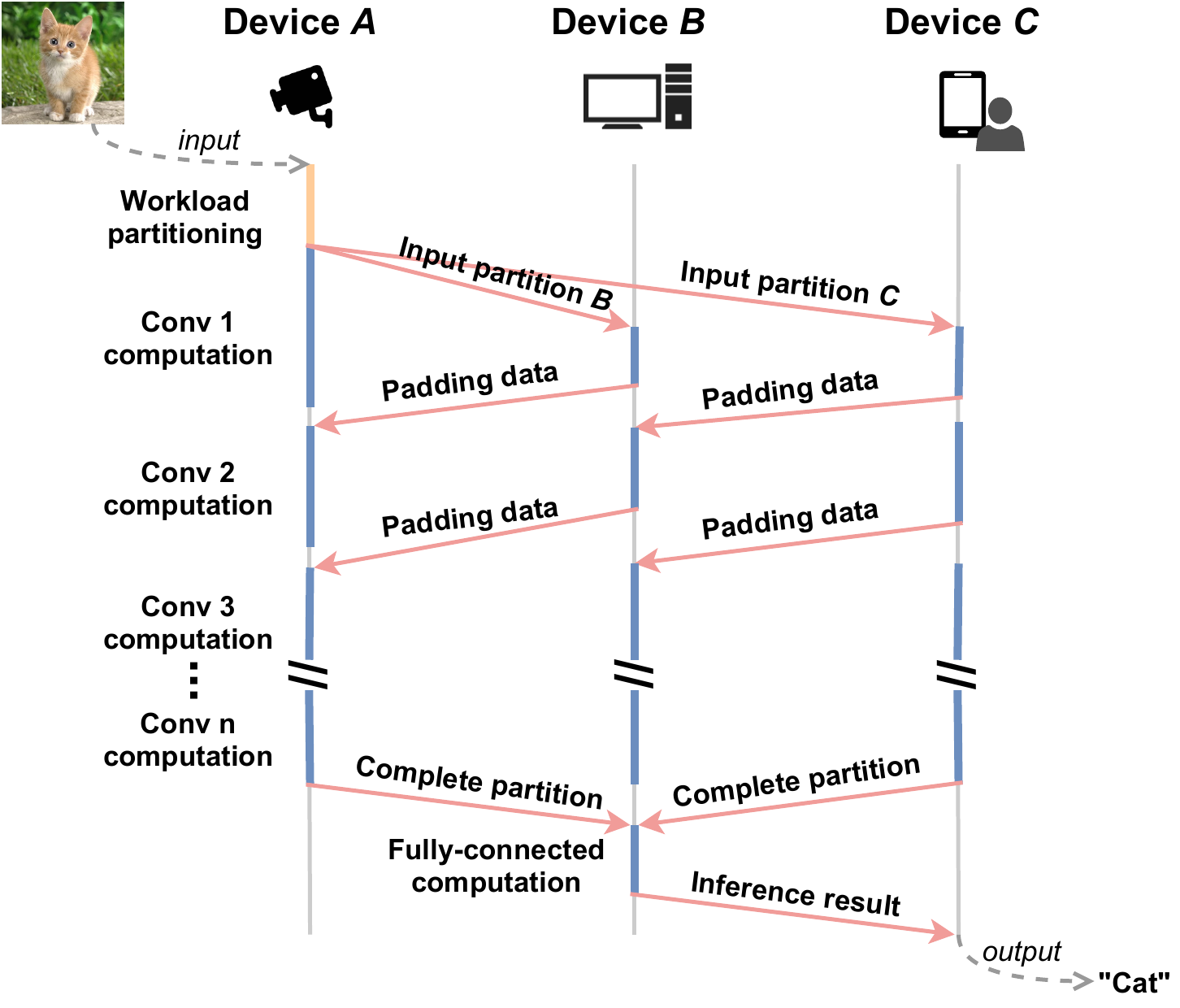}
    \caption{
    The communication pattern of a CoEdge runtime instance with three devices. For convolution computation, each device pulls the necessary padding data from its neighboring device. For fully-connected computation, the feature map partitions are aggregated. The final inference result is transferred to a user-specified location, i.e., device \textit{C} in the figure.
    }
    \label{fig:modeling_communication}
\end{figure}

To reduce the communication between devices, some prior works \cite{zhou2019adaptive, zhao2018deepthings} exploit sending redundant data in advance to avoid the padding issue.
However, while transferring redundant data takes additional communication cost, 
preparing necessary data beforehand for a number of CNN layers incurs extra storage overhead.
In this work, we address the padding issue by imposing a principle that requires the allocated partition size in the neighboring device to be not smaller than the padding size, unless it owns no partition.
This principle ensures that the padding data can be always acquired from only the neighboring device as long as it has data.
That is, the transmission of the padding data merely happens once, and thus we reduce the overhead in establishing additional connections.
To illustrate that, Fig. \ref{fig:modeling_communication} show the communication pattern of the example in Fig. \ref{fig:workflow}.
Initially, the input image is partitioned and distributed to corresponding devices, along with the padding data for the first convolution layer.
For the following layers, each device only connects to its neighbor device and fetches padding data for convolution computation.
This pattern holds until all convolutions are completed, and then the separated feature map partitions are aggregated to one of the devices for fully-connected computation.
The inference result is finally returned to a user-specified device (device \textit{C} for example in Fig. \ref{fig:modeling_communication}).

Under this principle, finding an appropriate workload assignment matters significantly for system performance.
For instance, offloading a large portion of workload to a device that owns high bandwidth but poor computing capability may not lead to a lower execution latency.
To deploy cooperative inference optimally, we need to match the assigned partition size to the computation and communication resource of each device.
We achieve this goal by designing a workload partitioning algorithm that is adaptive to the computing capabilities of available devices and the dynamic network status.

\section{Adaptive Workload Partitioning}
\label{sec:formulation}

The objective of optimizing workload allocation is to improve cooperative inference performance in both latency and energy metrics.
For the simplicity of problem definition, we target at meeting the latency requirement while minimizing the energy costs. 
Assuming an execution deadline $D$, the optimization problem is to optimally allocate the workload so that the system energy consumption is minimized while promising the execution deadline $D$, given the available computation and communication resources.

In the following, we present a detailed formulation of this problem and our workload partitioning algorithm.

\subsection{Problem Formulation}
\label{sec:modeling}

We assume that the devices are available and relatively stable during the inference runtime.
This can be relevant as executing an inference task is typically in a period of seconds, and many edge environments are maintained statically in independent spaces, such as smart home and smart factory.
Besides, the underlying support of intelligent services in such scenarios usually employs a few commonly-adopted DNN models and frequently run similar types of DNN inference tasks.
Therefore, we suppose that the DNN models have been loaded ahead of inference queries, and can be used to compute input tensors as soon as necessary data are prepared.

Since a CNN model typically encompasses many layers, we model the cooperative inference process from single-layer to multi-layer, progressively.
The single-layer formulation focuses on sketching the workload partitioning constraints and shaping the performance of single-layer, while the multi-layer part aims at summarizing the system behavior for the whole workflow.
Prior to that, we define the necessary concepts and notations as follows:

\begin{itemize}
    \item A layer $l$ is an algorithmic operation in a CNN model. In our formulation, a layer refers to either a convolution (Conv) or a fully-connected (FC) layer. Given a CNN model, $\mathcal{L} = [1,2, \cdots, L]$ denotes the layers in order.
    \item A partitioning solution $\pi$ is a group of coterminous partitions of the input image, which is generated by piece-wise partitioning along one dimension. 
    For the input partition assigned to device $i$, $a_i$ represents the number of rows that it covers.
    Hence, given the devices' indices $\mathcal{N} = [1, 2, \cdots, N]$, $\pi = [a_1, a_2, \cdots, a_N]$.
    We denote the workload as the input feature map partition to be processed on each DNN layer.
    For layer $l$, the workload size of the $i$-th partition is $r_{li}$, which can be obtained by calculating the partition's data size.
    \item A configuration tuple $(k, c_{\textbf{in}}, c_{\textbf{out}}, s, p)_{li}$ denotes the $l$-th layer's computation task on the $i$-th partition, which is characterized by the layer's configuration, i.e., convolution kernel size $k$, input channels $c_{\textbf{in}}$, output channels $c_{\textbf{out}}$, stride $s$, and padding $p$. 
    This tuple is applicable for both convolution and FC layers since FC computation can be viewed as a special case\footnote{This tuple depicts a fully-connected computation when the input feature map's size is $1 \times 1 \times c_{\textbf{in}}$, the output feature map's size is $1 \times 1 \times c_{\textbf{out}}$, kernel size $k=1$, stride $s=1$, and padding $p=0$.}.
    In particular, as discussed in Section \ref{sec:partition}, the padding size of convolutional layers is supposed to be smaller than the size of the partition on the last neighboring device, unless it owns no data.
    We formulate this principle as Eq. (\ref{eq:padding_constraint0}), where $\mathbbm{1}_{\{a_{i}>0\}}$ is an indicative function that values 1 if $a_{i}>0$ or else 0.
    This constraint is essentially equivalent to the disjunction of $a_{i} \geq p_{i+1}$ and $a_{i}=0$.
    \item A resource tuple $(\rho, f, m, P^c, P^x)_i$ specifics the resource profile of device $i$.
    Here, $\rho$ is defined as the computing intensity (in processing cycles per 1KB input) of the given DNN model, which is measured by application-driven offline profiling \cite{miettinen2010energy} in the setup phase.
    $f$ is the device's CPU frequency, reflecting its computing capability in a coarse granularity.
    $m$ is the available maximum memory capacity for inference tasks.
    For a single device that only processes CNN workloads, $m$ is the volume of memory excluding the space taken by the underlying system services, e.g., I/O services, compiler, etc.
    $P^c$ and $P^x$ denote the computation power and the wireless transmission power, respectively.
\end{itemize}

\subsubsection{Single-Layer Formulation}
\label{sec:single_layer_formulation}
There are some numerical constraints on the partition sizes.
Eq. (\ref{eq:padding_constraint0}) imposes the size restriction with the padding size as discussed in Section \ref{sec:partition}, and Eq. (\ref{eq:nature_constraint}) claims $a_i$ is a nonnegative integer.
Eq. (\ref{eq:sum_constraint}) presents that the concatenation size of all partitions along the height dimension equals to this dimension's size $H$.
The piece-wise partitioning can be conducted along either the height $H$ or width $W$ of the input.
In our experiments, we split along the height $H$ without loss of generality.
\begin{align}
    a_{i} \geq p_{i+1} \mathbbm{1}_{\{a_{i}> 0\}}, & \quad i\in\mathcal{N}, \label{eq:padding_constraint0} \\
    a_i \geq 0, a_i \in \mathbb{Z}, & \quad i\in\mathcal{N}, \label{eq:nature_constraint} \\
    \sum_{i\in\mathcal{N}} a_i = & H. \label{eq:sum_constraint}
\end{align}

The workload size $r_{li}$ of a partition is constrained by the device's available memory capacity $m_i$, as in Eq. (\ref{eq:memory_constraint}).
Here, we only limits the memory footprint on the size of per-layer inputs for the sake of simplicity, while the runtime memory may not be exactly $r_{li}$.
For practical deployment cases, emerging techniques on characterizing the detailed CNN execution memory (e.g., \cite{yao2018fastdeepiot}) can be adopted, and the deep learning platform-related memory footprint can be added into the left-hand side of Eq. (\ref{eq:memory_constraint}) as an enhancement. 
\begin{align}
    r_{li} \leq&  m_i, \quad i\in \mathcal{N}, l\in\mathcal{L}. \label{eq:memory_constraint}
\end{align}

During a single layer's execution, the system takes time and energy on two aspects, computation and communication.
For computation, we calculate the latency and energy by first approximating the computing cycles of given partitions.
As demonstrated in previous empirical studies \cite{miettinen2010energy, wen2012energy, cui2017software}, for many data processing tasks as exemplified by data encoding and decoding, the required computing cycles are proportional to their input data sizes.
This means that, a constant computing intensity (in computing cycles per unit data) exists for such tasks, and we can use it to capture the effective computing capability of a specific device.
Existing literature, such as \cite{mohammed2020distributed, mukherjee2020distributed, xu2020rjcc}, has leveraged this observation to characterize deep learning workloads, and in this work, we adopt it to estimate the computing cycles amount given the partitions and DNN layers.
Concretely, in Eq. (5), we assess the total processing cycles of the $i$-th partition by multiplying the device's computing intensity $\rho_i$ with the workload size $r_{li}$. 
Moreover, for each respective DNN layer, CNN inference typically conducts a feed-forward execution without any branch operation or recurrent computation \cite{sze2017efficient}, indicating that its execution latency is approximately linear to the computing cycles.
Therefore, the latency $T^c_{li}$ for computing layer $l$ is then appraised via dividing the total processing cycles by the computation frequency $f_i$, and the energy is the product of $T^c_{li}$ and the computation power $P^c_i$ in Eq. (\ref{eq:computation_energy}).
Note that Eq. (\ref{eq:computation_energy}) only reckons on dynamic energy.
Static energy consumption, e.g., those for maintaining basic system-level services, are not considered in our formulation.
\begin{align}
    T^c_{li} =& \frac{\rho_i r_{li}}{f_i}, \quad i\in\mathcal{N}, l\in\mathcal{L}, \label{eq:computation_time} \\
    E^c_{li} =& P^c_i T^c_{li}, \quad i\in\mathcal{N}, l\in\mathcal{L}. \label{eq:computation_energy}
\end{align}

For communication, let $b_{i,j}$ be the available bandwidth between devices $i$ and $j$. 
Particularly, $j=i$ implies delivering data from a device to itself, and $b_{i,i}$ is the memory bandwidth.
In our experiment, $b_{i,i}$ is set as 12.8GB/s by default, which is the typical memory bandwidth of DDR3 \cite{malladi2012towards}.
Initially, the communication occurs when the master device (noted as device $\textbf{M}$) distributes input partitions to worker devices, the transmission time is therefore calculated in the $l=1$ case of Eq. (\ref{eq:transmission_time}).
For communication of pulling the padding data from the neighboring device, its transmission time is described as the $l>1$ case.
For the sake of simplicity, Eq. (\ref{eq:transmission_time}) does not take the queuing delays into account since we are optimizing inference for respective single image input.
Streaming input, in which case the queuing delays significantly matter, are left for future work.
With the transmission power $P^x_i$, we acquire the dynamic energy of communicating with device $i$ on layer $l$ in Eq. (\ref{eq:transmission_energy}).
\begin{align}
    T^x_{li} &=
    \begin{cases}
    \frac{a_i}{b_{\textbf{M},i}}, & l = 1, i\in\mathcal{N},\\
    \frac{p_{li}}{b_{i,i+1}}, & l > 1, l\in\mathcal{L}, i\in\mathcal{N},
    \end{cases}
     \label{eq:transmission_time} \\
    E^x_{li} &= P^x_iT^{x}_{li}, \quad i\in\mathcal{N}, l\in\mathcal{L}. \label{eq:transmission_energy}
\end{align}

\subsubsection{Multi-Layer Formulation}
\label{sec:multi_layer_formulation}

\begin{figure}[htbp]
  \centering
  \includegraphics[width=0.9\linewidth]{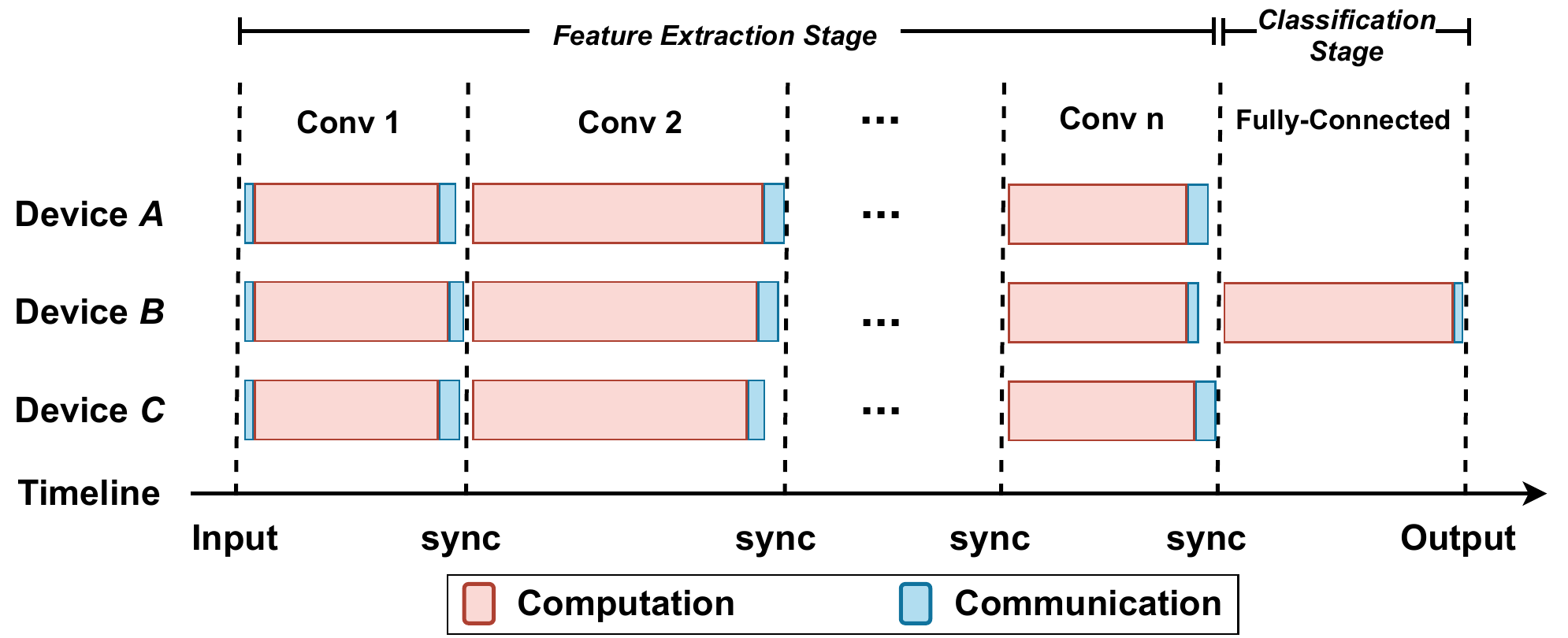}
  \caption{The job breakdown of a CoEdge runtime instance with three devices. Each device processes computation and communication jobs alternately, and the system performs in a Bulk Synchronous Parallel (BSP) mechanism.}
  \label{fig:lifetime}
\end{figure}

The key challenge of extending the formulation from a single layer to multiple layers lies in the synchronization mechanism during parallel processing.
Fig. \ref{fig:lifetime} presents a job breakdown of a CoEdge instance - processing one image over three devices.
As we can see, each device processes computation and communication jobs alternately, and they trigger synchronization periodically whenever a communication job (except for the initial communication job) is accomplished.
The contents of the synchronizations are the requisite padding data for convolutional computation.
During the interval between two synchronizations, there is no data dependency between devices, and thus they process jobs in parallel.
These scattered feature map partitions are finally aggregated at the classification stage for FC computation.
Hence, the whole process works in a Bulk Synchronous Parallel (BSP) mechanism \cite{valiant1990bridging}.

To summarize the cost of the whole process, we denote $E^c$ and $E^x$ as the total energy consumption of computation and communication, respectively, which are obtained by summing up the energy of all devices for all layers in Eq. (\ref{eq:energy_computation_all}) and (\ref{eq:energy_transmission_all}).
We count the total physical latency $T$ according to the BSP model and obtain Eq. (\ref{eq:time_all}).
Concretely, we acquire $T$ by calculating the maximum latency of all devices and then summing up the physical latency of all intervals.
It is worth noting that Eq. (\ref{eq:time_all}) has counted the latency of FC layers, as the maximum latency of all device is essentially that of the selected device in the classification stage.
\begin{align}
    E^c &= \sum_{l \in \mathcal{L}} \sum_{i \in \mathcal{N}} E^c_{li}, \label{eq:energy_computation_all} \\
    E^x &= \sum_{l \in \mathcal{L}} \sum_{i \in \mathcal{N}} E^x_{li}, \label{eq:energy_transmission_all} \\
    T &= \sum_{l \in \mathcal{L}} \max_{i \in \mathcal{N}} (T^c_{li}+T^x_{li}). \label{eq:time_all}
\end{align}

Given the execution deadline $D$, the targeted problem is to decide an optimal partitioning solution $\pi=[a_1, a_2, \cdots, a_N]$ with the objective of minimizing total energy without violating the execution deadline $D$.
Hence, we can formulate the cooperative inference optimization as the following problem.
\begin{align}
    \text{$\mathcal{P}$1 : min.} \quad & E^c + E^x \notag \\ 
    \text{s.t.} \quad & T \leq D,  \notag\\ 
    & (\ref{eq:padding_constraint0}), (\ref{eq:nature_constraint}), (\ref{eq:sum_constraint}), (\ref{eq:memory_constraint}). \notag 
\end{align}

\begin{theorem} \label{theorem:NP-hard}
Problem $\mathcal{P}$1 is an NP-hard problem.
\end{theorem}

We prove Theorem \ref{theorem:NP-hard} by identifying $\mathcal{P}$1 as an Integer Linear Programming problem (detailed in Appendix A).
In a typical smart factory deployment, there may be tens or even hundreds of edge devices, indicating that the decision space of $\mathcal{P}$1 can be huge (which grows exponentially with the increase of edge devices' amount) according to Theorem \ref{theorem:NP-hard}.
Therefore, to generate a solution in real-time, it is necessary to find an efficient solving method to $\mathcal{P}$1.

\subsection{Problem Transformation}

A Linear Programming (LP) problem is a kind of optimization towards a linear objective function subject to linear equality or inequality constraints, and the Integer Linear Programming (ILP) problem is a special case where all optimization variables are integers \cite{dantzig1998linear}.
As proved in Appendix A, $\mathcal{P}$1 is an ILP.
The difficulty of solving $\mathcal{P}$1 lies in the discreteness of integer variable $a_i$.
To produce a feasible solution to $\mathcal{P}$1 efficiently, we relax $\mathcal{P}$1 by introducing a continuous variable $\lambda_i$ to approximate $a_i$.
Eq. (\ref{eq:x2lambda}) defines the relation between $\lambda_i$ and $a_i$, where $H$ is the input's height and $\lambda_i$ describes the proportion that the $i$-th partition covers.
Since the input of CNN inference are usually of a large size (e.g., typically of 224 $\times$ 224 size from ImageNet \cite{deng2009imagenet} dataset), the approximation error is tiny and tolerated.
Eq. (\ref{eq:lambda_indicative}), (\ref{eq:lambda_range}), and (\ref{eq:lambda_sum})  show the numerical constraints for $\lambda_i$, which are derived from Eq. (\ref{eq:padding_constraint0}), (\ref{eq:nature_constraint}), and (\ref{eq:sum_constraint}), respectively.
\begin{align}
    a_i = \lambda_i & H, \quad i\in\mathcal{N}, \label{eq:x2lambda} \\
    \lambda_{i} H \geq p_{i+1} & \mathbbm{1}_{\{\lambda_{i}>0\}}, \quad i\in\mathcal{N}, \label{eq:lambda_indicative}\\ 
    \lambda_i \ge & 0 , \quad i\in\mathcal{N}, \label{eq:lambda_range} \\
    \sum_{i \in \mathcal{N}} & \lambda_i = 1.  \label{eq:lambda_sum}
\end{align}

Eq. (\ref{eq:lambda_indicative}) is essentially equivalent to the expression of $\lambda_i H \geq p_{i+1}$ or $\lambda_i = 0$.
Since $\lambda_i = 0$ is a potential solution, it is feasible to separate solving $\lambda_i$'s value and checking whether $\lambda_i \geq p_{i+1}$ to two steps.
Therefore, we relax the constraint Eq. (\ref{eq:lambda_indicative}) as $\lambda_i H \geq 0$, i.e., $\lambda_i \geq 0$, and $\mathcal{P}$1 can be transformed into the following problem $\mathcal{P}$2.
\begin{align}
    \text{$\mathcal{P}$2 : min.} \quad & E^c + E^x \notag \\
    % \text{s.t.} \quad & T \leq D, \notag \\ 
    %     & r_{li} \leq m_i, \quad i\in\mathcal{N}, l \in \mathcal{L}, \notag \\ 
    %     & \lambda_{i} \geq 0, \quad i\in\mathcal{N}, \notag \\ 
    %     & \sum_{i \in \mathcal{N} } \lambda_i = 1. \notag  
    \text{s.t.} \quad & T \leq D, \notag \\ 
        & (\ref{eq:memory_constraint}), (\ref{eq:lambda_range}), (\ref{eq:lambda_sum}). \notag
\end{align}

$\mathcal{P}$2 is fundamentally a special case of $\mathcal{P}$1.
Particularly, on the solution to $\mathcal{P}$2, there may be some devices that are assigned with tiny workload ($\exists i \in \mathcal{N}, 0 \leq \lambda_i < p_{i+1}$), while on the solution to $\mathcal{P}$1, the workload size on all devices must be larger than or equal to the padding data size unless it is zero ($\forall i \in \mathcal{N}, \lambda_i \geq p_{i+1}$ or $\lambda_i = 0$).
Regardless of the potential solution $\lambda_i = 0$ to problem $\mathcal{P}$1, the main difference between $\mathcal{P}$1 and $\mathcal{P}$2 is the setting of threshold, i.e., $\mathcal{P}$1 sets the threshold as $p_{i+1}$ while $\mathcal{P}$2 sets $0$.
Hence, we can exploit $\mathcal{P}$2's solution to iteratively approach $\mathcal{P}$1's solution by checking whether it satisfies the threshold constraint Eq. (\ref{eq:lambda_indicative}).

\begin{theorem} \label{theorem:convex}
Problem $\mathcal{P}$2 is a Linear Programming problem.
\end{theorem}

Theorem \ref{theorem:convex} (proved in Appendix B) reveals that $\mathcal{P}$2 is a LP, which can be efficiently solved by existing mature programming solvers (e.g., CPLEX \cite{cplex2009v12}). 
By them, it is feasible to fast approximate the solution to $\mathcal{P}$1.

\subsection{Workload Partitioning Algorithm Design}
\label{sec:algorithm}

We propose a threshold-based workload partition algorithm for $\mathcal{P}$1 using existing programming solvers, as presented in Algorithm \ref{algo:partition}.
The key idea of Algorithm \ref{algo:partition} is to gradually narrow down the selection of participating devices (i.e., the devices that are assigned with workload) by checking the threshold constraint and iteratively approach the solution.

The input of Algorithm \ref{algo:partition} includes CNN layer configurations, available computation and communication resources, and the execution deadline.
These inputs provide parameters for $\mathcal{P}$1 and $\mathcal{P}$2.
The output is the workload partitioning solution to $\mathcal{P}$1.

Algorithm \ref{algo:partition} begins with checking whether $\mathcal{N}$ is empty - if $\mathcal{N}$ is an empty set, there is no available devices to perform cooperative inference, and thus no feasible solution to $\mathcal{P}$1.
Otherwise, we solve $\pi$ from $\mathcal{P}$2 with a LP solver.
Then we check whether the obtained $\pi$ satisfy Eq. ($\ref{eq:lambda_indicative}$), the threshold constraint of $\mathcal{P}$1.
If so, the current version of $\pi$ is a feasible solution and is immediately returned.
Or else, there must be some elements in $\pi$ that are smaller than the required padding size.
In this case, we remove part of these unsatisfied elements from the available devices list: 
firstly we remove zero elements since the zero workload assignment indicates that the device would not participate in cooperative inference; 
next we find the minimum from the rest elements in $\pi$ and remove it from $\mathcal{N}$.
After that, the algorithm goes to the next recursion to acquire the new partition solution with the updated $\mathcal{N}$ and checks the result for $\mathcal{P}$1 again.
The recursion continues until it finds a feasible solution - if any - or a null flag as the unfeasible signal (this could happen when the deadline constraint is too strict to satisfy).

The programming solver for our LP problem runs efficiently. 
In our experiment we use CPLEX and the runtime overhead is smaller than 1ms. 
Since the total recursion times in Algorithm \ref{algo:partition} will not exceed $N$ (the total number of available devices), the solving process of Algorithm \ref{algo:partition} is very fast ($<$10 ms) and will not cause side-effect on the pursuit of latency SLO.

\begin{algorithm}[t] 
\caption{Workload Partitioning Algorithm} 
\label{algo:partition} 
\begin{algorithmic}[1] 
\REQUIRE ~~\\ 
$\mathcal{N}$: Available devices $[1,2,\cdots,N]$ \\
$\mathcal{L}$: CNN layers $[1,2,\cdots,L]$ \\
$(k, c_{\textbf{in}}, c_{\textbf{out}}, s, p)_{li}, \forall i \in \mathcal{N}, \forall l \in \mathcal{L} $: Configuration tuples \\
$(\rho, f, m, P^c, P^x)_i, \forall i \in \mathcal{N} $: Resources tuples \\
$b_{i,j}, \forall i,j \in \mathcal{N} $: Bandwidths \\
$D$: Execution deadline \\
\ENSURE ~~\\ 
$\pi$: Assigned workload proportions $[\lambda_1, \lambda_2, \cdots, \lambda_N]$ 

\STATE \textbf{Procedure} \textsc{Partition}($\mathcal{N}$)
\STATE \quad \textbf{if} $\mathcal{N}$ is empty \textbf{then}
\STATE \quad \quad \textbf{return} NULL \quad \quad \quad \quad \quad \quad $\rhd$ {no feasible solution} 
\STATE \quad Solve $\pi$ from $\mathcal{P}$2
\STATE \quad \textbf{if} $\pi$ satisfy Eq. (\ref{eq:lambda_indicative}) \textbf{then}
\STATE \quad \quad \textbf{return} $\pi$
\STATE \quad \textbf{else}
\STATE \quad \quad Find the index set $\mathcal{N}_0$ of zero elements in $\pi$
\STATE \quad \quad Find the minimum element $\lambda_{m}$ in $\pi$
\STATE \quad \quad $\mathcal{N} \leftarrow \mathcal{N} - \mathcal{N}_0 - \{m\}$
\STATE \quad \quad \textbf{return} \textsc{Partition}($\mathcal{N}$)
\STATE \textbf{end Procedure}
\end{algorithmic}
\end{algorithm}

\section{Prototype Implementation}
\label{sec:implementation}

We employ TensorFlow Lite \cite{tensorflow} as the backend engine to execute CNN layers, and implement the communication module based on gRPC \cite{gRPC2019}.
In the following, we provide the implementation details of CoEdge.

\textbf{Deployment and profiling.} Since any one of the devices in the environment may launch a CNN inference task, the employed CNN models are trained and installed on all devices in advance.
As the model is installed, we use TensorFlow benchmark tool to profile the latency of one inference and measure the energy with the Monsoon High Voltage Power Monitor \cite{monsoon}.
For each CNN model, we run it for once as warm-up and then record the execution time with 50 runs without break.
The aim of warm-up running is to alleviate the impact of weight loading and TensorFlow initiation since we have omitted these overheads in the formulation.
The execution tasks on all devices are the same - perform CNN inference on the same image from ImageNet \cite{deng2009imagenet}.
We take the mean values as the measuring results and derive the resource tuple parameters based on them.

The computation frequency $f$ is directly from known specifications.
With $f$ and the measured latency, we can estimate the total computing cycles of one inference.
Dividing the cycles' amount to the processed image size yields the computing intensity $\rho$.
We obtain the memory capacity $m$ by observing the available memory space of an idle system.
For power parameters $P^c$ and $P^x$, we measure them by calculating the measured computation/communication energy and delay.

\textbf{Workload partitioning and distribution.}
To create the workload allocation plan efficiently, We run the workload partitioning algorithm based on IBM ILOG CPLEX \cite{cplex2009v12}, a linear programming solver package.
If the algorithm returns a feasible solution, we segment the input image accordingly and send the partitions to the corresponding devices.
Otherwise, the algorithm returns an infeasible signal, which means the deadline is set too strict.
In this case, we choose to offload all workload to the device that can minimize the end-to-end execution latency.

\begin{figure}[t]
    \centering
    \includegraphics[width=0.7\linewidth]{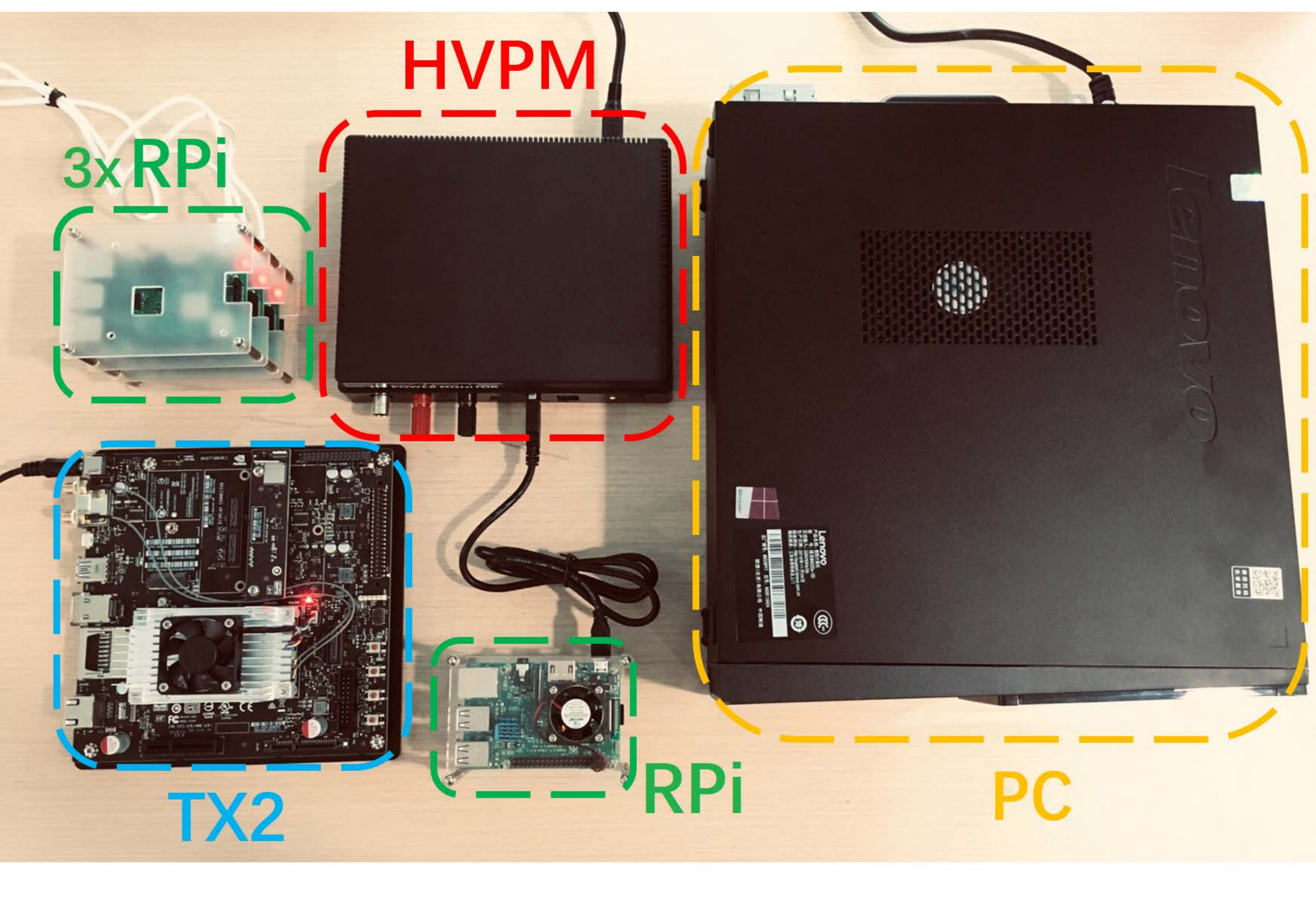}
    \caption{Our experimental prototype uses four Raspberry Pis (RPi), one Jetson TX2 and one desktop PC. Their specifications are listed in Table \ref{tab:pi}, \ref{tab:jetson} and \ref{tab:PC}. We employ the Monsoon High Voltage Power Monitor (HVPM) to measure the energy.}
    \label{fig:hardware}
\end{figure}

\textbf{Runtime communication.} During the runtime, each device needs to fetch the padding data from its neighboring device. 
Due to the limited computing capability, a device may be still working on generating the output feature map partition when a padding pulling request arrives.
To accommodate this case, we block the pulling request until the needed data is prepared.
Note that such circumstance is rare since our workload partitioning algorithm has optimized the workload allocation to match devices' computing capabilities.
Under this plan, the execution time on each device is reasonably close and the utilization of computing resources are maximized as much as possible.
Moreover, our workload partitioning algorithm supposes that participated devices can well communicate with each other during the runtime.
However, the devices can accidentally break down or temporarily unavailable in real-world deployment.
This raises robustness issues, which be discussed in Section \ref{sec:discussion}.

\section{Performance Evaluation}
\label{sec:evaluation}

In this section, we evaluate the performance of CoEdge prototype in terms of inference latency and dynamic energy.
We also explore the impact of deadline setting, the system scalability and the adaptability to network fluctuation.

\subsection{Experimental Setup}

\begin{table}[t]
\caption{Desktop PC Specifications}
\label{tab:PC}
\centering
\begin{tabular}{|c|c|c|}
\hline
Hardware & \multicolumn{2}{c|}{Specifications} \\ \hline \hline
CPU      & \multicolumn{2}{c|}{3.60GHz 8-Core Intel i7-7700} \\ \hline
Memory   & \multicolumn{2}{c|}{2666MHz 16GB DDR4} \\ \hline
GPU      & \multicolumn{2}{c|}{GeForce GTX 1050 (Pascal) 640 CUDA core} \\ \hline \hline
Power    & \begin{tabular}[c]{@{}c@{}}Idle\\ CPU Fully Loaded\\ GPU Fully Loaded\end{tabular} & \begin{tabular}[c]{@{}c@{}}80W\\ 180W\\ 200W\end{tabular} \\ \hline
\end{tabular}
\end{table}

\begin{table}[]
\caption{ Inference Latency (ms) And Computation \protect\\ Intensity (cycles/KB) of Basic Implementation}
\label{tab:computation_intensity}
\centering
\begin{tabular}{|c|c|c|c|c|c|c|}
\hline
\multirow{2}{*}{Model} & \multicolumn{2}{c|}{Raspberry Pi} & \multicolumn{2}{c|}{Jetson TX2} & \multicolumn{2}{c|}{Desktop PC} \\ \cline{2-7} 
          & Lat. & Inten. & Lat. & Inten. & Lat. & Inten. \\ \hline \hline
AlexNet   & 302  & 615  & 89   & 301  & 46   & 282  \\ \hline
VGG-f     & 276  & 563  & 83   & 283  & 44   & 269  \\ \hline
GoogLeNet & 769  & 1568 & 227  & 772  & 114  & 698  \\ \hline
MobileNet & 226  & 461  & 71   & 239  & 37   & 226  \\ \hline
\end{tabular}
\end{table}

\textbf{Prototype.} we implement CoEdge prototype with six devices: four Raspberry Pi 3, one Jetson TX2, and one desktop PC, as shown in Fig. \ref{fig:hardware}.
The Raspberry Pi 3 and the Jetson TX2 represent weak IoT devices and mobile AI platforms.
Besides, we take a desktop PC to emulate small edge servers.
The specifications of the three types of devices are provided in Table \ref{tab:pi}, \ref{tab:jetson} and \ref{tab:PC}.
We employ the Monsoon High Voltage Power Monitor (HVPM) \cite{monsoon} to measure the energy.
For bandwidth control, We use the traffic control tool tc \cite{tc2019}, which is able to limit the bandwidth under the setting value.

\textbf{Workload.} In our prototype, we use TensorFlow Lite \cite{tensorflow} to implement four typical CNN models: AlexNet\cite{krizhevsky2012imagenet}, VGG-f \cite{simonyan2014very}, GoogLeNet \cite{szegedy2015going}, and MobileNet \cite{howard2017mobilenets}, all of which are trained before deployment.
Table \ref{tab:computation_intensity} presents the reported latency of basic implementation and the computing intensity on different platforms.
We set the workload as the image classification task on one ImageNet \cite{deng2009imagenet} image.
The average inference latency and computation intensity of one hundred runs are taken as the results.
During the runtime we turn off all applications except for necessary OS background services.

\textbf{Approaches.} We compare CoEdge with the following relative approaches. 
(1) \textit{MoDNN} \cite{mao2017modnn} adopts the same piece-wise partitioning mechanism as CoEdge, but decides partition sizes in proportion to the devices' computing capabilities without considering network conditions. 
(2) \textit{Musical Chair} \cite{hadidi2018distributed} is a cooperative inference system that exploits both data and model parallelism. 
For each layer, it chooses one of the parallelisms and accordingly partitions the workloads in equal proportion.
(3) \textit{Local} approach executes CNN inference at the master device solely. In our experiment, the local approach is the baseline, and we fix the master device as a certain Raspberry Pi 3.

\subsection{Performance Comparison}

\begin{figure}[t]
    \centering
    \includegraphics[width=\linewidth]{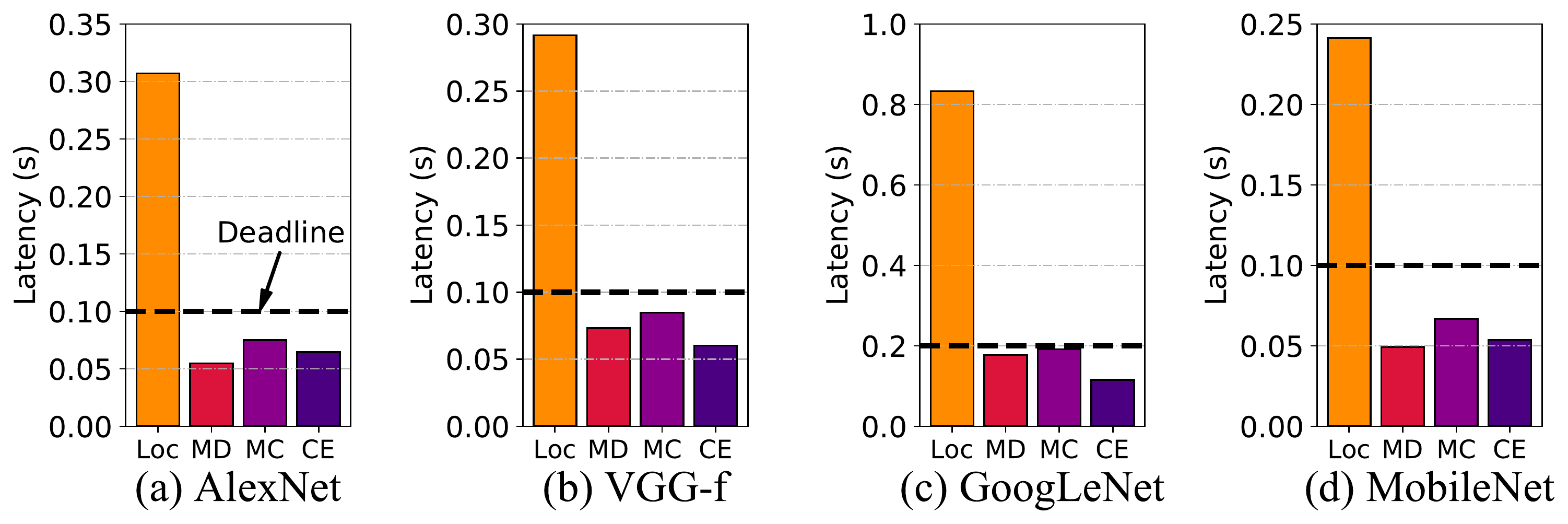}
    \caption{The end-to-end latency of different approaches running four DNN models. The deadline of AlexNet, VGG-f, GoogLeNet, and MobileNet are set as 100ms, 100ms, 200ms, and 100ms, respectively. }
    \label{fig:evaluation_latency}
\end{figure}

\begin{figure}[t]
    \centering
    \includegraphics[width=\linewidth]{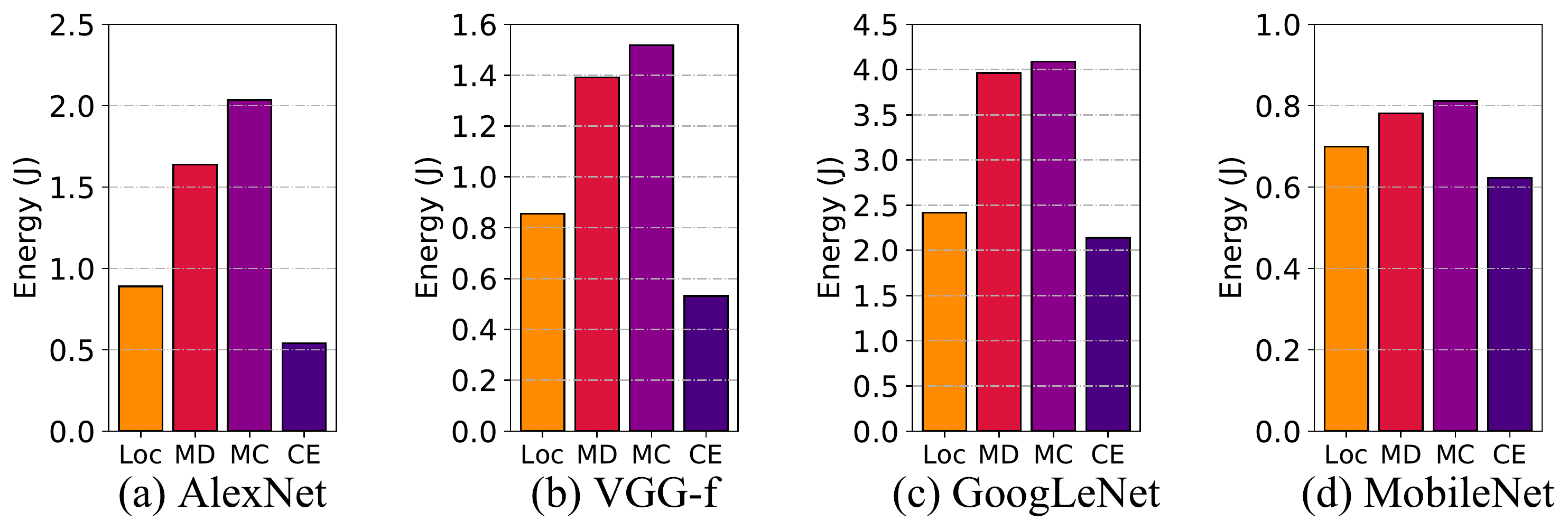}
    \caption{The dynamic energy consumption of different approaches running four DNN models. We use the testbed in Fig. \ref{fig:hardware} that consists of four Raspberry Pi 3, one Jetson TX2, and one desktop PC. All experimental settings are the same as that in the experiment of Fig. \ref{fig:evaluation_latency}.}
    \label{fig:evaluation_energy}
\end{figure}

Fig. \ref{fig:evaluation_latency} and Fig. \ref{fig:evaluation_energy} show the latency and dynamic energy results of different models with the local approach (Loc), MoDNN (MD), Musical Chair (MC), and CoEdge (CE).
The results in these two figures are measured at the same experimental settings, and the maximum bandwidth between devices are fixed at 1MB/s.
We set the deadline for executing the four models as 100ms, 100ms, 200ms, and 100ms, respectively, marked as dashed lines in Fig. \ref{fig:evaluation_latency}.

As shown in Fig. \ref{fig:evaluation_latency}, CoEdge, Musical Chair and MoDNN always accomplish inference within the deadline.
As the most time-consuming option, the local approach is the only one that violates the latency requirement, and CoEdge achieves 7.21$\times$$\sim$4.49$\times$ latency speedup over it.
Comparing the local approach with the other ones reflects the power of cooperative inference gained by harvesting vicinal edge resources.
Among the three cooperative approaches, Musical Chair takes higher latency that the other two.
This is because that Musical Chair directly split the workload in equal proportion ignoring the resources heterogeneity.
CoEdge and MoDNN perform closely in the latency metric, but differs their energy costs in the energy metrics.

As an evidence, Fig. \ref{fig:evaluation_energy} shows the least dynamic energy consumption that CoEdge takes comparing with other approaches.
CoEdge saves up to 66.9\%,  64.9\%, 46.0\%, and 25.5\% energy for four models, respectively (comparing with Muscial Chair).
To the baseline (local approach), CoEdge saves 39.2\%, 37.8\%, 11.5\%, and 10.9\% energy.
CoEdge saves energy prominently for AlexNet and VGG-f, but promote not so much for GoogLeNet and MobileNet.
This attributes to the structure of CNN models.
GoogLeNet's completed block structure comprises a crowd of layers, which incurs frequent data exchanges in cooperative inference.
At the opposite end of the spectrum, MobileNet uses a simplified structure and has been well optimized for local inference in embedded devices, which limits the improvement space for cooperative inference.
It is worth noting that the local approach consumes less energy than MoDNN and Musical Chair.
The reasons come from two aspects.
On the one hand, the local approach does not incur communication costs, while MoDNN and Musical Chair need frequent cross-device communication during the runtime, which takes energy.
On the other hand, the optimization of MoDNN and Musical Chair does not consider the power characteristics of different types of devices so that the workload are processed in an energy-lavish manner.
In contrast, by jointly optimizing the computation-communication tradeoff provided devices' computing capabilities and network conditions, CoEdge achieves the lowest energy costs.

\subsection{Performance under Varying Deadlines}

\begin{figure}[t]
\centering
\begin{minipage}[t]{0.47\linewidth}
\centering
\includegraphics[width=0.85\linewidth]{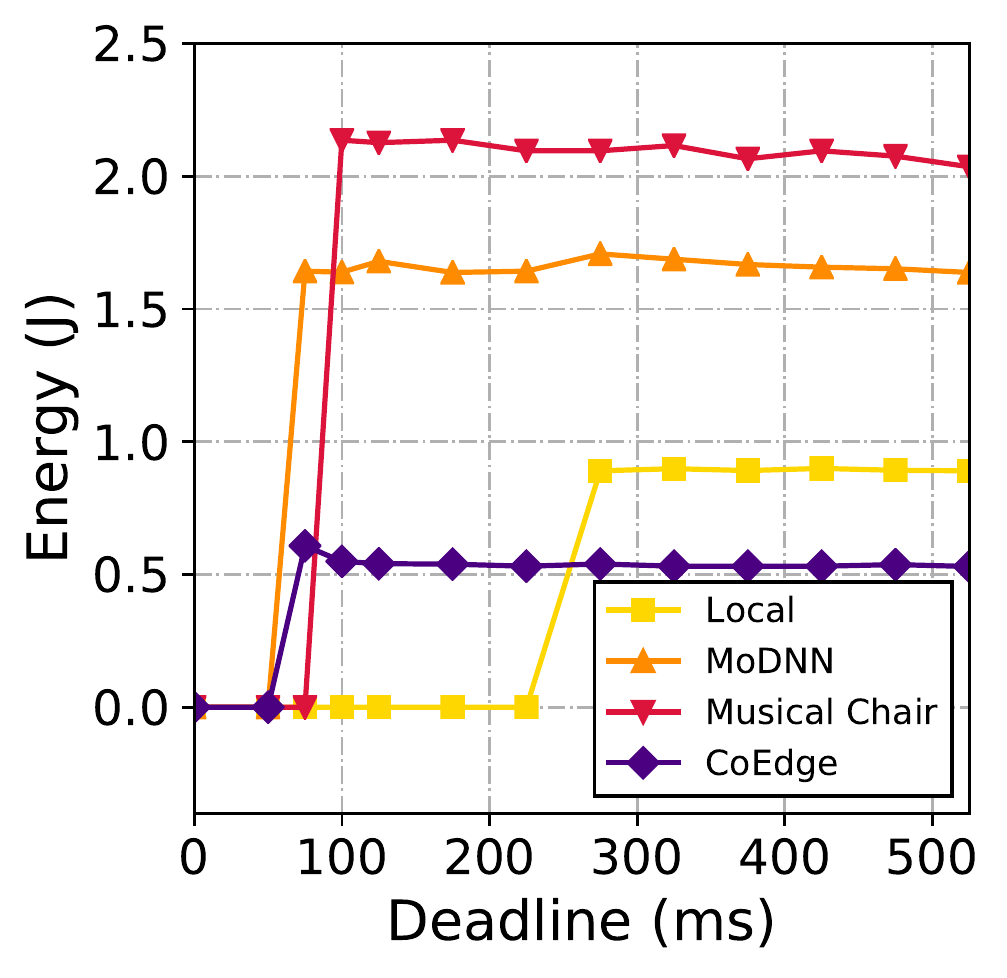}
\caption{The dynamic energy consumption of four approaches under varying deadlines built on the testbed in Fig. \ref{fig:hardware}. The result is recorded as zero if the approach fails to finish inference within the deadline.}
\label{fig:evaluation_deadline_energy}
\end{minipage}
\quad
\begin{minipage}[t]{0.47\linewidth}
\centering
\includegraphics[width=\linewidth]{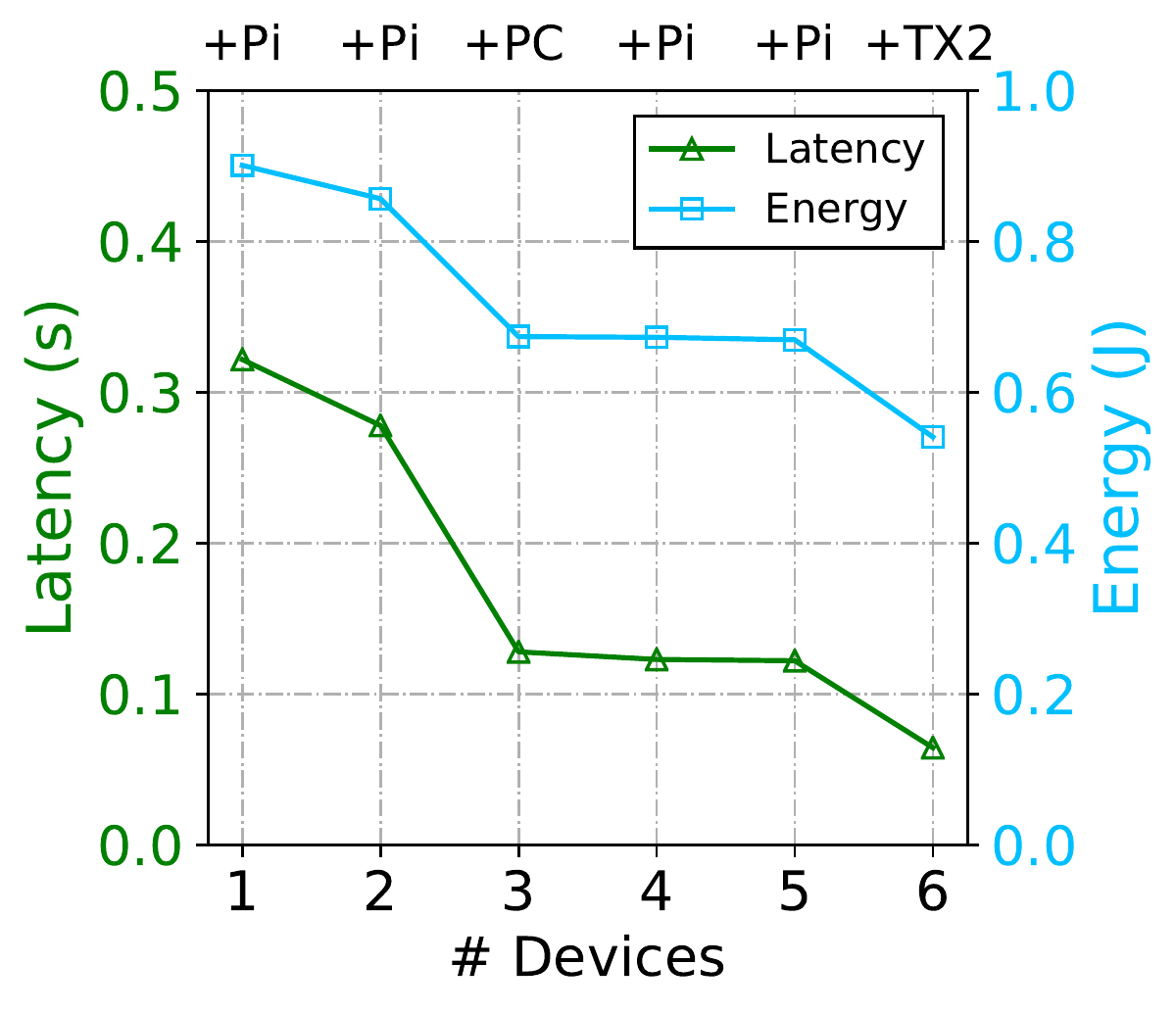}
\caption{The latency and dynamic energy results of CoEdge with varying number of devices. The top text indicates which type of device are newly added to the cluster.}
\label{fig:evaluation_scalability}
\end{minipage}
\end{figure}

In this experiment, we explore how the deadline setting impacts the system performance.
We run AlexNet to process one image input.
The bandwidths between devices are fixed as 1MB/s.
Fig. \ref{fig:evaluation_deadline_energy} shows the dynamic energy results as a function of deadlines.
To emphasize the deadline constraint, we plot the energy result as zero if it fails to accomplish the inference within the deadline.
When the latency requirement is very stringent ($\leq$ 50ms), all approaches miss the deadline.
At the 75ms deadline, CoEdge and MoDNN first succeed completing the execution, while CoEdge takes lower energy costs than MoDNN.
When the deadline sets 100ms, Musical Chair finishes a full inference, but it consumes higher energy than MoDNN and CoEdge.
As the latency requirement gradually relaxes, the local approach finally achieves.

Note that CoEdge shows a converged declining curve as the deadline postpones.
CoEdge takes higher energy under a stringent deadline (75ms) than a loose deadline (500ms).
This is because CoEdge's optimization puts latency constraint in prior to energy optimization.
In the case with a very strict latency requirement, CoEdge prefers to sacrifice some energy-saving to latency reduction.
With the requirement loosens, the pressure of satisfying deadline constraint gradually relaxes and CoEdge will transfer emphasis on energy optimization.
When the deadline is adequately slack, it is not a constraint to our optimization anymore, in which case the change of that will not impact the workload allocation plan of CoEdge and thus the dynamic energy result keeps stable.

\subsection{Scalability}

To evaluate CoEdge's scalability, we measure the latency and energy by incrementally adding devices to the experimental cluster.
We fix the bandwidth as 1MB/s and set a loose deadline of 500ms.
The inference task is run AlexNet with one image for classification.
We add devices in the following order: Raspberry Pi, Raspberry Pi, desktop PC,  Raspberry Pi, Raspberry Pi, and Jetson TX2.
Fig. \ref{fig:evaluation_scalability} presents the measuring results of CoEdge, where the top text shows the adding devices orderly.
With the increase of the cluster scale, both the latency and dynamic energy drop.
In particular, there is a distinctive decrease when adding PC (2$\rightarrow$3) and Jetson TX2 (5$\rightarrow$6).
This is reasonable as the cluster adds a relatively much more powerful device (PC or Jetson) comparing with the Pi at these two points.
When the scale is extended to 4 or 5, the latency and dynamic energy keeps approximately stable, indicating that CoEdge runs almost invariant solutions.
This is because CoEdge prefers allocating a major portion of the workload to the devices with higher computing capabilities, e.g., the PC.
When the cluster already owns such powerful device, adding a weak one (e.g., Pi) will not make distinct changes to the workload allocation plan, and thus the system performance keeps steady as the previous organization.

\subsection{Adaptability to Network Fluctuation}
\begin{figure}[t]
    \centering
    \includegraphics[width=0.95\linewidth]{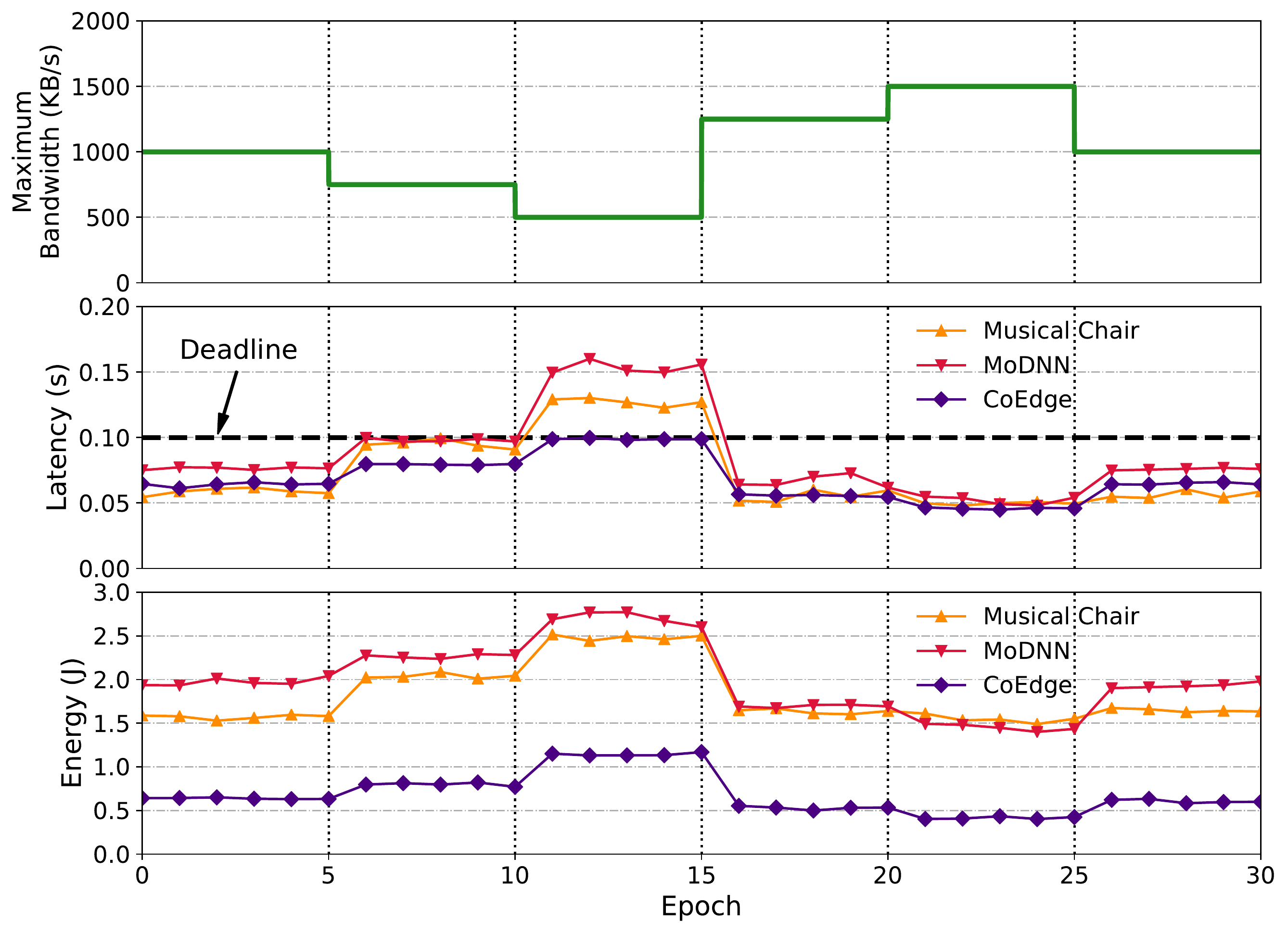}
    \caption{The latency and dynamic energy results under varying network status. The deadline for CoEdge is set as 100ms.}
    \label{fig:evaluation_interference}
\end{figure}

In this experiment, we record the system performance of different cooperative approaches with varying bandwidths.
We run AlexNet with one image on the six-device cluster, and the deadline is 100ms, plotted as the dashed line in Fig. \ref{fig:evaluation_interference}.
For each epoch, CoEdge captures the available bandwidths and triggers a reprogramming on workload partitioning if the bandwidths change.
This reprogramming process takes tiny overhead, reporting less than 10ms in the experiment.
We adjust the bandwidth settings between devices in different periods.
The top subfigure in Fig. \ref{fig:evaluation_interference} presents the network fluctuation with the bandwidths of 1000KB/s, 750KB/s, 500KB/s, 1250KB/s, 1500KB/s, and 1000KB/s, respectively.
As the bandwidth changes, all three approaches vary their performance.
The performance variation comes from two reasons.
On the one hand, the communication overhead for necessary data exchange during cooperative inference depends on the network conditions.
On the other hand, for CoEdge, diverse bandwidths yield diverse partitioning plans and therefore impact the performance of cooperation.
In most cases, the latency results of the approaches are approximate.
On the energy side, however, CoEdge outperforms other approaches all the time.
In particular, when the bandwidth drops at 500KB/s (Epoch 11-15), only CoEdge's execution satisfies the deadline.
In other epochs, as the bandwidth becomes higher, CoEdge adjusts the workload partitioning and the energy costs decrease.

\section{Related Work}
\label{sec:related_work}

Previous research efforts in enabling artificial intelligence at the network edge can be divided into three directions: cloud-assisted execution, local resource exploitation, and multi-device collaboration.

\textbf{Cloud-assisted execution.}
Cloud-assisted approaches offload DNN inference workload from local to the cloud fully or partially \cite{han2016mcdnn, li2019edge, kang2017neurosurgeon, zeng2019boomerang, teerapittayanon2017distributed, jeong2018ionn}.
MCDNN \cite{han2016mcdnn} fully offloads DNN computation.
It creates DNN model variants and selects one from them to maximize the accuracy under resource constraints, while CoEdge does not involve accuracy issues as the model and data are never modified.
Neurosurgeon \cite{kang2017neurosurgeon} proposes a partially offloading solution, which decides an intermediate partition point in the DNN structure to keep front layers locally and offload rear layers to the cloud.
DDNN \cite{teerapittayanon2017distributed} leverages a similar principle and partitions DNN layers in a cloud-edge-device hierarchy.
However, it requests to retrain the DNN model in scheduling, while CoEdge does not require any retraining work.
The cloud-assisted approaches have been widely-adopted in mobile scenarios, e.g., drone-enabled vehicles tracking \cite{ballotta2020computation}, robotics-based vision applications \cite{chinchali2018neural, chinchali2019network, tanwani2020rilaas}.
On these specific cases, the cloud's functionality is further optimized to adjust demands.
For example, RILaaS \cite{tanwani2020rilaas} introduces a Robot-Inference-and-Learning-as-a-Service platform with robotics-oriented features such as reliable network protocol, secure authentication, and REST front-end API.

\textbf{Local resource exploitation.}
Local approaches keep all computation locally and optimize the performance through hardware specialization or model modification \cite{kim2019mulayer, han2015deep, yao2018fastdeepiot, xu2019reform, oh2018portable}.
Hardware specialization generally centers around basic DNN operations
(e.g., matrix multiplication, convolution) to develop efficient
hardware accelerators, e.g., ARM ML Processor \cite{armML20}, Google Edge TPU \cite{edgeTPU20}.
Some other works target to optimize the utilization of existing hardware.
For example, $\mu$Layer \cite{kim2019mulayer} accelerates inference in layer granularity by simultaneously utilizing heterogeneous processors inside an edge device.
Model modification typically uses model compression technique, e.g., model sparsification and quantization \cite{han2015deep}.
ReForm \cite{xu2019reform} reconfigurates CNN models by model pruning and selective computing to reduce inference latency on mobile devices.
On the same goal, libnumber \cite{oh2018portable} employs quantization technique to optimize number representation in low-level, reducing both model size and inference latency.
CoEdge is orthogonal to such optimizations since it does not apply any structural modifications to the employed DNN.

\textbf{Multi-device collaboration.}
Multi-device approaches executes DNN inference using a cluster of devices in the edge environment \cite{zhao2018deepthings, hu2020fast, mao2017modnn}.
Within this category, previous works optimize workload distribution in two ways, layer fusion and workload size adjustment.
Layer fusion partitions the feature maps in a fixed pattern, and distributes workload with redundant data - the padding data - to avoid data requests between devices during the runtime.
% In cooperative inference, the redundant data is the padding part between two neighboring partitions.
Under this mechanism, DeepThings \cite{zhao2018deepthings} fuses front convolutional layers and parallelizes these layers on multiple devices.
The follow-up work \cite{zhou2019adaptive} generalizes the fusion operation to all layers and takes resources heterogeneity into account.
It designs a dynamic-programming-based fusion searching strategy to adaptively decide which layers are fused and which layers are directly parallelized.
Workload size adjustment accommodates the workload allocation to minimize the end-to-end inference latency.
MoDNN \cite{mao2017modnn} segments workload greedily and assigns more workload to the devices with higher computing capability without considering the network conditions.
Musical Chair \cite{hadidi2018distributed} introduces a partitioning algorithm integrating data and model parallelism, and partitions the feature maps in equal proportion.
Based on Musical Chair, the subsequent work \cite{hadidi2020towards, cao2019edge, hadidi2019robustly} improves distributed CNN execution in terms of latency performance, scalability, and robustness, respectively.
% For example, to improve system scalability, virtualization techniques are employed to containerize and manage CNN computation tasks on edge devices \cite{cao2019edge}.

Our work falls into the mutli-device collaboration category, and combines the parallel workflow of layer-fusion techniques \cite{zhao2018deepthings, zhou2019adaptive} and the partitioning mechanism of workload adjustment approaches \cite{hadidi2018distributed, mao2017modnn}.
Beyond combining the novel designs of these two lines, CoEdge jointly considers available computation and communication resources and improves the workload allocation on heterogeneous devices via an adaptive algorithm, which has not been addressed in prior works.

\section{Discussion and Future directions}
\label{sec:discussion}

In this section, we discuss the limitation and extension of CoEdge, and provide some future research directions.

\textbf{Robustness and Generalization.} 
As a distributed system, crash of any participant or network timeout can result in a system breakdown of cooperative inference.
To increase the robustness against such faults, it may be helpful to design modularity \cite{chinchali2019network} for the system or reserve intermediate result backups periodically.
Another direction is to further generalize and optimize the system workflow for more sophisticated model structures.
Only applying workload partitioning over the whole network may not well fit more complicated architectures as the feature maps of the deeper layers usually exhibit in a smaller height and width.

\textbf{Other optimizing objectives.}
CoEdge focuses on optimizing the dynamic energy consumption with preset deadlines for CNN inference.
Modifying the objective function of the constrained programming can steer CoEdge to meet other priorities.
For example, one can adopt a performance preference by setting a tunable-weighted synthesis of latency and energy.
Alternatively, taking static energy consumption into account may produce a more energy-friendly workload allocation plan.
Another potential objective is accuracy.
Although CoEdge does not sacrifice any accuracy theoretically, running DNNs on some minitype edge devices may still loss precision owing to the limitations of their modest computing capability and the execution mechanism of DNN frameworks.
Characterizing and optimizing such accuracy issue is practically significant for edge deployment.

\textbf{Utilizing edge-oriented resources.}
Recent technical progresses on edge computing enhancement in computation (e.g., pluggable Google Edge TPU \cite{edgeTPU20}, Intel Movidius Neural Compute Stick \cite{movidius20}) could potentially benefit CoEdge performance.
For example, by equipping a Raspberry Pi with an Edge TPU, CoEdge may choose to remain the input workload mainly or even completely in situ.
This requires more efforts on shaping and utilizing the emerging elastic computing resources.
Moreover, improvements on the communication side, e.g., 5G and mmWave, can also boost cooperative edge intelligence.

\section{Conclusion}
\label{sec:concolusion}

In this paper, we present CoEdge, a distributed DNN computing system that orchestrates cooperative DNN inference over heterogeneous edge devices.
We explore the workflow of cooperative inference and formulate it as a constrained optimization problem, which is NP-hard.
To solve it efficiently, we design a workload partitioning algorithm to decide efficient partitioning policy in real-time.
By jointly optimizing computation and communication, CoEdge can find the optimal workload partitioning plan that minimizes the system energy cost while promising execution latency requirements.
Experimental evaluations using a realistic prototype show 7.21$\times$$\sim$4.49$\times$ latency speedup over the local approach and up to 25.5\%$\sim$66.9\% energy saving comparing with existing approaches for four widely-adopted DNN models.

\ifCLASSOPTIONcaptionsoff
  \newpage
\fi

\bibliographystyle{IEEEtran}
\bibliography{main.bib}

\begin{thebibliography}{10}
\providecommand{\url}[1]{#1}
\csname url@samestyle\endcsname
\providecommand{\newblock}{\relax}
\providecommand{\bibinfo}[2]{#2}
\providecommand{\BIBentrySTDinterwordspacing}{\spaceskip=0pt\relax}
\providecommand{\BIBentryALTinterwordstretchfactor}{4}
\providecommand{\BIBentryALTinterwordspacing}{\spaceskip=\fontdimen2\font plus
\BIBentryALTinterwordstretchfactor\fontdimen3\font minus
  \fontdimen4\font\relax}
\providecommand{\BIBforeignlanguage}[2]{{%
\expandafter\ifx\csname l@#1\endcsname\relax
\typeout{** WARNING: IEEEtran.bst: No hyphenation pattern has been}%
\typeout{** loaded for the language `#1'. Using the pattern for}%
\typeout{** the default language instead.}%
\else
\language=\csname l@#1\endcsname
\fi
#2}}
\providecommand{\BIBdecl}{\relax}
\BIBdecl

\bibitem{stojkoska2017review}
B.~L.~R. Stojkoska and K.~V. Trivodaliev, ``A review of internet of things for
  smart home: Challenges and solutions,'' \emph{Journal of Cleaner Production},
  vol. 140, pp. 1454--1464, 2017.

\bibitem{shrouf2014smart}
F.~Shrouf, J.~Ordieres, and G.~Miragliotta, ``Smart factories in industry 4.0:
  A review of the concept and of energy management approached in production
  based on the internet of things paradigm,'' in \emph{2014 IEEE international
  conference on industrial engineering and engineering management}.\hskip 1em
  plus 0.5em minus 0.4em\relax IEEE, 2014, pp. 697--701.

\bibitem{gerla2014internet}
M.~Gerla, E.-K. Lee, G.~Pau, and U.~Lee, ``Internet of vehicles: From
  intelligent grid to autonomous cars and vehicular clouds,'' in \emph{2014
  IEEE world forum on internet of things (WF-IoT)}.\hskip 1em plus 0.5em minus
  0.4em\relax IEEE, 2014, pp. 241--246.

\bibitem{rahmani2015smart}
A.-M. Rahmani, N.~K. Thanigaivelan, T.~N. Gia, J.~Granados, B.~Negash,
  P.~Liljeberg, and H.~Tenhunen, ``Smart e-health gateway: Bringing
  intelligence to internet-of-things based ubiquitous healthcare systems,'' in
  \emph{2015 12th Annual IEEE Consumer Communications and Networking Conference
  (CCNC)}.\hskip 1em plus 0.5em minus 0.4em\relax IEEE, 2015, pp. 826--834.

\bibitem{shi2020carpool}
Q.~Shi and X.~Chen, ``Carpool for big data: Enabling efficient crowd
  cooperation in data market for pervasive ai,'' \emph{IEEE Transactions on
  Vehicular Technology}, 2020.

\bibitem{acharya2017deep}
U.~R. Acharya, S.~L. Oh, Y.~Hagiwara, J.~H. Tan, M.~Adam, A.~Gertych, and
  R.~San~Tan, ``A deep convolutional neural network model to classify
  heartbeats,'' \emph{Computers in biology and medicine}, vol.~89, pp.
  389--396, 2017.

\bibitem{sze2017efficient}
V.~Sze, Y.-H. Chen, T.-J. Yang, and J.~S. Emer, ``Efficient processing of deep
  neural networks: A tutorial and survey,'' \emph{Proceedings of the IEEE},
  vol. 105, no.~12, pp. 2295--2329, 2017.

\bibitem{deng2013new}
L.~Deng, G.~Hinton, and B.~Kingsbury, ``New types of deep neural network
  learning for speech recognition and related applications: An overview,'' in
  \emph{2013 IEEE International Conference on Acoustics, Speech and Signal
  Processing}.\hskip 1em plus 0.5em minus 0.4em\relax IEEE, 2013, pp.
  8599--8603.

\bibitem{simonyan2014very}
K.~Simonyan and A.~Zisserman, ``Very deep convolutional networks for
  large-scale image recognition,'' \emph{arXiv preprint arXiv:1409.1556}, 2014.

\bibitem{li2019edge}
E.~Li, L.~Zeng, Z.~Zhou, and X.~Chen, ``Edge ai: On-demand accelerating deep
  neural network inference via edge computing,'' \emph{IEEE Transactions on
  Wireless Communications}, vol.~19, no.~1, pp. 447--457, 2019.

\bibitem{ouyang2018follow}
T.~Ouyang, Z.~Zhou, and X.~Chen, ``Follow me at the edge: Mobility-aware
  dynamic service placement for mobile edge computing,'' \emph{IEEE Journal on
  Selected Areas in Communications}, vol.~36, no.~10, pp. 2333--2345, 2018.

\bibitem{chen2018thriftyedge}
X.~Chen, Q.~Shi, L.~Yang, and J.~Xu, ``Thriftyedge: Resource-efficient edge
  computing for intelligent iot applications,'' \emph{IEEE network}, vol.~32,
  no.~1, pp. 61--65, 2018.

\bibitem{zhou2019paving}
Z.~Zhou, X.~Chen, E.~Li, L.~Zeng, K.~Luo, and J.~Zhang, ``Edge intelligence:
  Paving the last mile of artificialintelligence with edge computing,''
  \emph{Proceedings of the IEEE}, 2019.

\bibitem{zhao2018deepthings}
Z.~Zhao, K.~M. Barijough, and A.~Gerstlauer, ``Deepthings: Distributed adaptive
  deep learning inference on resource-constrained iot edge clusters,''
  \emph{IEEE Transactions on Computer-Aided Design of Integrated Circuits and
  Systems}, vol.~37, no.~11, pp. 2348--2359, 2018.

\bibitem{pi2019}
R.~P. Foundation, ``Raspberry pi 3,''
  \url{https://www.raspberrypi.org/products/raspberry-pi3-model-b/}, accessed
  December 15, 2019.

\bibitem{jetson2019}
NVIDIA, ``Nvidia jetson tx,''
  \url{https://developer.nvidia.com/embedded/jetson-tx2}, accessed December 15,
  2019.

\bibitem{monsoon}
Monsoon, ``High voltage power monitor,''
  \url{https://www.msoon.com/high-voltage-power-monitor}, accessed May 25,
  2019.

\bibitem{hadidi2018distributed}
R.~Hadidi, J.~Cao, M.~Woodward, M.~S. Ryoo, and H.~Kim, ``Distributed
  perception by collaborative robots,'' \emph{IEEE Robotics and Automation
  Letters}, vol.~3, no.~4, pp. 3709--3716, 2018.

\bibitem{krizhevsky2012imagenet}
A.~Krizhevsky, I.~Sutskever, and G.~E. Hinton, ``Imagenet classification with
  deep convolutional neural networks,'' in \emph{Advances in neural information
  processing systems}, 2012, pp. 1097--1105.

\bibitem{tensorflow}
``Tensorflow benchmark tool,'' \url{
  https://github.com/tensorflow/tensorflow/tree/r1.4/tensorflow/tools/benchmark},
  accessed May 15, 2019.

\bibitem{deng2009imagenet}
J.~Deng, W.~Dong, R.~Socher, L.-J. Li, K.~Li, and L.~Fei-Fei, ``Imagenet: A
  large-scale hierarchical image database,'' in \emph{2009 IEEE conference on
  computer vision and pattern recognition}.\hskip 1em plus 0.5em minus
  0.4em\relax Ieee, 2009, pp. 248--255.

\bibitem{tc2019}
T.~L. Foundation, ``Tc – show / manipulate traffic control settings,''
  \url{https://www.linux.com/tutorials/tc-show-manipulate-traffic-control-settings/},
  accessed December 15, 2019.

\bibitem{lin2013network}
M.~Lin, Q.~Chen, and S.~Yan, ``Network in network,'' \emph{International
  Conference on Learning Representations}, 2014.

\bibitem{he2016deep}
K.~He, X.~Zhang, S.~Ren, and J.~Sun, ``Deep residual learning for image
  recognition,'' in \emph{Proceedings of the IEEE conference on computer vision
  and pattern recognition}, 2016, pp. 770--778.

\bibitem{zhou2019adaptive}
L.~Zhou, M.~H. Samavatian, A.~Bacha, S.~Majumdar, and R.~Teodorescu, ``Adaptive
  parallel execution of deep neural networks on heterogeneous edge devices,''
  in \emph{Proceedings of the 4th ACM/IEEE Symposium on Edge Computing}, 2019,
  pp. 195--208.

\bibitem{miettinen2010energy}
A.~P. Miettinen and J.~K. Nurminen, ``Energy efficiency of mobile clients in
  cloud computing,'' \emph{HotCloud}, vol.~10, no. 4-4, p.~19, 2010.

\bibitem{yao2018fastdeepiot}
S.~Yao, Y.~Zhao, H.~Shao, S.~Liu, D.~Liu, L.~Su, and T.~Abdelzaher,
  ``Fastdeepiot: Towards understanding and optimizing neural network execution
  time on mobile and embedded devices,'' in \emph{Proceedings of the 16th ACM
  Conference on Embedded Networked Sensor Systems}.\hskip 1em plus 0.5em minus
  0.4em\relax ACM, 2018, pp. 278--291.

\bibitem{wen2012energy}
Y.~Wen, W.~Zhang, and H.~Luo, ``Energy-optimal mobile application execution:
  Taming resource-poor mobile devices with cloud clones,'' in \emph{2012
  Proceedings Ieee Infocom}.\hskip 1em plus 0.5em minus 0.4em\relax IEEE, 2012,
  pp. 2716--2720.

\bibitem{cui2017software}
Y.~Cui, J.~Song, K.~Ren, M.~Li, Z.~Li, Q.~Ren, and Y.~Zhang, ``Software defined
  cooperative offloading for mobile cloudlets,'' \emph{IEEE/ACM Transactions on
  Networking}, vol.~25, no.~3, pp. 1746--1760, 2017.

\bibitem{mohammed2020distributed}
T.~Mohammed, C.~Joe-Wong, R.~Babbar, and M.~Di~Francesco, ``Distributed
  inference acceleration with adaptive dnn partitioning and offloading,'' in
  \emph{IEEE INFOCOM 2020-IEEE Conference on Computer Communications}.\hskip
  1em plus 0.5em minus 0.4em\relax IEEE, 2020, pp. 854--863.

\bibitem{mukherjee2020distributed}
M.~Mukherjee, V.~Kumar, A.~Lat, M.~Guo, R.~Matam, and Y.~Lv, ``Distributed deep
  learning-based task offloading for uav-enabled mobile edge computing,'' in
  \emph{IEEE INFOCOM 2020-IEEE Conference on Computer Communications Workshops
  (INFOCOM WKSHPS)}.\hskip 1em plus 0.5em minus 0.4em\relax IEEE, 2020, pp.
  1208--1212.

\bibitem{xu2020rjcc}
S.~Xu, Q.~Liu, B.~Gong, F.~Qi, S.~Guo, X.~Qiu, and C.~Yang, ``Rjcc:
  Reinforcement learning based joint communicational-and-computational resource
  allocation mechanism for smart city iot,'' \emph{IEEE Internet of Things
  Journal}, 2020.

\bibitem{malladi2012towards}
K.~T. Malladi, F.~A. Nothaft, K.~Periyathambi, B.~C. Lee, C.~Kozyrakis, and
  M.~Horowitz, ``Towards energy-proportional datacenter memory with mobile
  dram,'' in \emph{2012 39th Annual International Symposium on Computer
  Architecture (ISCA)}.\hskip 1em plus 0.5em minus 0.4em\relax IEEE, 2012, pp.
  37--48.

\bibitem{valiant1990bridging}
L.~G. Valiant, ``A bridging model for parallel computation,''
  \emph{Communications of the ACM}, vol.~33, no.~8, pp. 103--111, 1990.

\bibitem{dantzig1998linear}
G.~B. Dantzig, \emph{Linear programming and extensions}.\hskip 1em plus 0.5em
  minus 0.4em\relax Princeton university press, 1998, vol.~48.

\bibitem{cplex2009v12}
I.~I. CPLEX, ``V12. 1: User’s manual for cplex,'' \emph{International
  Business Machines Corporation}, vol.~46, no.~53, p. 157, 2009.

\bibitem{gRPC2019}
Google, ``gprc - a rpc library and framework,'' \url{https://grpc.io}, accessed
  December 15, 2019.

\bibitem{szegedy2015going}
C.~Szegedy, W.~Liu, Y.~Jia, P.~Sermanet, S.~Reed, D.~Anguelov, D.~Erhan,
  V.~Vanhoucke, and A.~Rabinovich, ``Going deeper with convolutions,'' in
  \emph{Proceedings of the IEEE conference on computer vision and pattern
  recognition}, 2015, pp. 1--9.

\bibitem{howard2017mobilenets}
A.~G. Howard, M.~Zhu, B.~Chen, D.~Kalenichenko, W.~Wang, T.~Weyand,
  M.~Andreetto, and H.~Adam, ``Mobilenets: Efficient convolutional neural
  networks for mobile vision applications,'' \emph{arXiv preprint
  arXiv:1704.04861}, 2017.

\bibitem{mao2017modnn}
J.~Mao, X.~Chen, K.~W. Nixon, C.~Krieger, and Y.~Chen, ``Modnn: Local
  distributed mobile computing system for deep neural network,'' in
  \emph{Design, Automation \& Test in Europe Conference \& Exhibition (DATE),
  2017}.\hskip 1em plus 0.5em minus 0.4em\relax IEEE, 2017, pp. 1396--1401.

\bibitem{han2016mcdnn}
S.~Han, H.~Shen, M.~Philipose, S.~Agarwal, A.~Wolman, and A.~Krishnamurthy,
  ``Mcdnn: An approximation-based execution framework for deep stream
  processing under resource constraints,'' in \emph{Proceedings of the 14th
  Annual International Conference on Mobile Systems, Applications, and
  Services}.\hskip 1em plus 0.5em minus 0.4em\relax ACM, 2016, pp. 123--136.

\bibitem{kang2017neurosurgeon}
Y.~Kang, J.~Hauswald, C.~Gao, A.~Rovinski, T.~Mudge, J.~Mars, and L.~Tang,
  ``Neurosurgeon: Collaborative intelligence between the cloud and mobile
  edge,'' in \emph{ACM SIGARCH Computer Architecture News}, vol.~45,
  no.~1.\hskip 1em plus 0.5em minus 0.4em\relax ACM, 2017, pp. 615--629.

\bibitem{zeng2019boomerang}
L.~Zeng, E.~Li, Z.~Zhou, and X.~Chen, ``Boomerang: On-demand cooperative deep
  neural network inference for edge intelligence on industrial internet of
  things,'' \emph{IEEE Network}, 2019.

\bibitem{teerapittayanon2017distributed}
S.~Teerapittayanon, B.~McDanel, and H.-T. Kung, ``Distributed deep neural
  networks over the cloud, the edge and end devices,'' in \emph{2017 IEEE 37th
  International Conference on Distributed Computing Systems (ICDCS)}.\hskip 1em
  plus 0.5em minus 0.4em\relax IEEE, 2017, pp. 328--339.

\bibitem{jeong2018ionn}
H.-J. Jeong, H.-J. Lee, C.~H. Shin, and S.-M. Moon, ``Ionn: Incremental
  offloading of neural network computations from mobile devices to edge
  servers,'' in \emph{Proceedings of the ACM Symposium on Cloud Computing},
  2018, pp. 401--411.

\bibitem{ballotta2020computation}
L.~Ballotta, L.~Schenato, and L.~Carlone, ``Computation-communication
  trade-offs and sensor selection in real-time estimation for processing
  networks,'' \emph{IEEE Transactions on Network Science and Engineering},
  2020.

\bibitem{chinchali2018neural}
S.~P. Chinchali, E.~Cidon, E.~Pergament, T.~Chu, and S.~Katti, ``Neural
  networks meet physical networks: Distributed inference between edge devices
  and the cloud,'' in \emph{Proceedings of the 17th ACM Workshop on Hot Topics
  in Networks}, 2018, pp. 50--56.

\bibitem{chinchali2019network}
S.~Chinchali, A.~Sharma, J.~Harrison, A.~Elhafsi, D.~Kang, E.~Pergament,
  E.~Cidon, S.~Katti, and M.~Pavone, ``Network offloading policies for cloud
  robotics: a learning-based approach,'' in \emph{Robotics: Science and
  Systems}, 2019, pp. 1--10.

\bibitem{tanwani2020rilaas}
A.~K. Tanwani, R.~Anand, J.~E. Gonzalez, and K.~Goldberg, ``Rilaas: Robot
  inference and learning as a service,'' \emph{IEEE Robotics and Automation
  Letters}, 2020.

\bibitem{kim2019mulayer}
Y.~Kim, J.~Kim, D.~Chae, D.~Kim, and J.~Kim, ``$\mu$layer: Low latency
  on-device inference using cooperative single-layer acceleration and
  processor-friendly quantization,'' in \emph{Proceedings of the Fourteenth
  EuroSys Conference 2019}, 2019, pp. 1--15.

\bibitem{han2015deep}
S.~Han, H.~Mao, and W.~J. Dally, ``Deep compression: Compressing deep neural
  networks with pruning, trained quantization and huffman coding,'' \emph{arXiv
  preprint arXiv:1510.00149}, 2015.

\bibitem{xu2019reform}
Z.~Xu, F.~Yu, C.~Liu, and X.~Chen, ``Reform: Static and dynamic resource-aware
  dnn reconfiguration framework for mobile device,'' in \emph{Proceedings of
  the 56th Annual Design Automation Conference 2019}, 2019, pp. 1--6.

\bibitem{oh2018portable}
Y.~H. Oh, Q.~Quan, D.~Kim, S.~Kim, J.~Heo, S.~Jung, J.~Jang, and J.~W. Lee, ``A
  portable, automatic data qantizer for deep neural networks,'' in
  \emph{Proceedings of the 27th International Conference on Parallel
  Architectures and Compilation Techniques}, 2018, pp. 1--14.

\bibitem{armML20}
ARM, ``Arm ml processor,''
  \url{https://www.arm.com/products/silicon-ip-cpu/ethos/ethos-n77}, accessed
  July 16, 2020.

\bibitem{edgeTPU20}
Google, ``Google edge tpu,'' \url{https://cloud.google.com/edge-tpu}, accessed
  July 16, 2020.

\bibitem{hu2020fast}
D.~Hu and B.~Krishnamachari, ``Fast and accurate streaming cnn inference via
  communication compression on the edge,'' in \emph{2020 IEEE/ACM Fifth
  International Conference on Internet-of-Things Design and Implementation
  (IoTDI)}.\hskip 1em plus 0.5em minus 0.4em\relax IEEE, 2020, pp. 157--163.

\bibitem{hadidi2020towards}
R.~Hadidi, J.~Cao, M.~S. Ryoo, and H.~Kim, ``Towards collaborative inferencing
  of deep neural networks on internet of things devices,'' \emph{IEEE Internet
  of Things Journal}, 2020.

\bibitem{cao2019edge}
J.~Cao, F.~Wu, R.~Hadidi, L.~Liu, T.~Krishna, M.~S. Ryoo, and H.~Kim, ``An
  edge-centric scalable intelligent framework to collaboratively execute dnn,''
  in \emph{Demo for SysML Conference, Palo Alto, CA}, 2019.

\bibitem{hadidi2019robustly}
R.~Hadidi, J.~Cao, M.~S. Ryoo, and H.~Kim, ``Robustly executing dnns in iot
  systems using coded distributed computing,'' in \emph{Proceedings of the 56th
  Annual Design Automation Conference 2019}, 2019, pp. 1--2.

\bibitem{movidius20}
Intel, ``Intel movidius neural compute stick,''
  \url{https://software.intel.com/content/www/us/en/develop/hardware/neural-compute-stick.html},
  accessed July 16, 2020.

\end{thebibliography}

\appendices
\section{Proof of Theorem 1}
\label{sec:proof1}
\begin{proof}
We reduce $P||C_{max}$ problem to a special case of $\mathcal{P}$1, where all the power parameters are set as 1. 
Since $P||C_{max}$ problem is NP-hard, $\mathcal{P}$1 is at least as hard as $P||C_{max}$ problem.

Firstly, we identify $\mathcal{P}$1 as an integer linear programming problem.
For the optimizing variable $a_i$, the constaints (\ref{eq:padding_constraint0}), (\ref{eq:nature_constraint}), and (\ref{eq:sum_constraint}) limit it into a range of nonnegative integers.
Using $a_i$, we can obtain the initial workload on each device by multiplying $a_i$ and the data size of each row.
Since the input feature maps of each layer are the output of the prior layer, we can derive the workload of each layer based on its specific configuration.
For example, for convolution operation, given the input feature map partition of size $(H,W,C_{\textbf{in}})$ (Height, Width, Channels) and the convolution kernel $(k,C_{\textbf{in}}, C_{\textbf{out}},s,p)$, the output size is $(\frac{H-k+2p}{s}+1, \frac{W-k+2p}{s}, C_{\textbf{out}})$.
Therefore, we can express $r_{li}$ linearly using $a_i$.
So do for $T^c_{li}$, $T^x_{li}, E^c_{li}, E^x_{li}$ according to Eq. (\ref{eq:computation_time}), (\ref{eq:transmission_time}), (\ref{eq:energy_computation_all}), and (\ref{eq:energy_transmission_all}).

For the deadline constraint $T = \sum_{l \in \mathcal{L}} \max_{i \in \mathcal{N}} (T^c_{li}+T^x_{li}) \leq D$, we transform it into a series of inequalities.
Assuming a sub-deadline $D_l$ for processing layer $l$, we have $\max_{i \in \mathcal{N}} (T^c_{li}+T^x_{li}) \leq D_l$, which is equivalent to $T^c_{l1}+T^x_{l1} \leq D_l, T^c_{l2}+T^x_{l2} \leq D_l, \cdots, T^c_{lN}+T^x_{lN} \leq D_l$.
Without loss of generality, we conduct this transformation to all interval and obtain $N\cdot L$ inequalities in total, i.e., $T^c_{li}+T^x_{li} \leq D_l, \forall_{i\in\mathcal{N}}, \forall_{l\in\mathcal{L}}$.
Given that $T^c_{li}$ and $T^x_{li}$ is linear with $a_i$, these inequalities are linear.

In conclusion, all the expressions in $\mathcal{P}$1 are either linear function or integer constraint, indicating that $\mathcal{P}$1 is an integer linear programming problem. 
Let the variables $a_i$ be the jobs to schedule and all power parameters be 1, we can reduce $P||C_{max}$ problem to $\mathcal{P}$1 by recognizing the total energy in $\mathcal{P}$1 as the processing time in $P||C_{max}$.
Since $P||C_{max}$ problem is NP-hard, $\mathcal{P}$1 is NP-hard.
\end{proof}

\section{Proof of Theorem 2}
\label{sec:proof2}
\begin{proof}
As we have discussed in the proof of Theorem \ref{theorem:NP-hard}, the objective function, the memory constraint, and the deadline constraint are linear with $a_i$.
In $\mathcal{P}$2, we substitute $\lambda_i H $ for $a_i$, therefore, the linear relationship is still satisfied and the variable is now continuous.
For the remaining numerical constraints $\lambda_i \geq 0$ and $\sum_{i \in \mathcal{N}} \lambda_i = 1$, they are still linear expressions.
In summary, all the functions in $\mathcal{P}$2 are linear toward the elements in $\pi$.
Hence, $\mathcal{P}$2 is a linear programming problem.
\end{proof}

\end{document}